\documentclass[10pt,aps,prd,twocolumn,showpacs,superscriptaddress,eqsecnum,longbibliography,nofootinbib]{revtex4-2}
\usepackage{mathrsfs,amsmath,amsthm,latexsym,amssymb,amsfonts,epsfig,cancel,enumerate,graphicx,subfigure,txfonts}
\usepackage{xcolor}
\definecolor{navy}{RGB}{0,0,150}
\usepackage[colorlinks, linkcolor=navy, citecolor=navy, urlcolor=navy, plainpages=false, pdfstartview=FitH]{hyperref}
\usepackage{appendix}
\allowdisplaybreaks
\newcommand{\GZU}{School of Physics, Guizhou University, Guiyang 550025, China}

\begin{document}
\title{Shadows of rotating black holes in effective quantum gravity}

\author{Zhenglong Ban}
\email{gs.zlban22@gzu.edu.cn }
\affiliation{\GZU}

\author{Jiawei Chen}
\email{gs.chenjw23@gzu.edu.cn}
\affiliation{\GZU}

\author{Jinsong Yang}
\email{jsyang@gzu.edu.cn}
\affiliation{\GZU}

\begin{abstract}
	
Recently, two new spherically symmetric black hole models with covariance have been proposed in effective quantum gravity. Based on these models, we use the modified Newman-Janis algorithm to generate two rotating quantum-corrected black hole solutions, characterized by three parameters, the mass $M$, the spin $a$, and the quantum parameter $\zeta$. To understand the effects of the quantum parameter $\zeta$ on these two rotating black holes, we investigate in detail the horizons and static limit surfaces. By constraining the possible range of the parameters, we study the shadows cast by these rotating black holes. The results indicate that for both rotating BHs, the parameter $\zeta$ mainly affects the shadow size in the non-extremal case, while it deforms the shadow shape by arising a cuspy edge in the near-extremal case.
\end{abstract}

\maketitle

\section{Introduction}

In recent years, the exploration of black holes (BHs) has been at the forefront of research in the field of theoretical physics. With breakthroughs in gravitational wave astronomy \cite{ LIGOScientific:2016aoc} and the rapid development of BH imaging technology \cite{Mann:2018xkm,EventHorizonTelescope:2019dse}, we have gained a deeper understanding of these enigmatic celestial objects in the universe. Due to the fundamental conflict between the singularity theorems and quantum mechanics, it is widely believed that general relativity (GR) may not fully describe gravitational phenomena across all scales in the universe. Therefore, one aspires to solve this problem within the framework of quantum gravity theory. In the context of quantum gravity, the study on BH not only advances our understanding of the nature of BH but also provides a possible pathway to test the theory of quantum gravity. Many influential quantum gravity theories have been proposed, among which loop quantum gravity (LQG) has received extensive attention and in-depth research \cite{Rovelli:2004tv,Thiemann:2007pyv,Ashtekar:2004eh,Han:2005km,Ashtekar:2005qt,Modesto:2008im,Perez:2017cmj,Ashtekar:2018lag,Bodendorfer:2019cyv,Kelly:2020uwj,Gan:2020dkb,Sartini:2020ycs,Song:2020arr,Zhang:2020qxw,Zhang:2021wex,Lewandowski:2022zce,Zhang:2023yps}. Recently, two new effective BH models motivated by LQG have been proposed in \cite{Zhang:2024khj}, which retain the general covariance of spacetime within the framework of the Hamiltonian approach to quantum gravity. General covariance is one of the core principles of GR, requiring that physical laws take the same form in all reference frames, thereby ensuring the consistency and universality of the theory. The models not only provide a platform for studying quantum gravity effects but also allow us to explore how these effects manifest without violating the fundamental principles of GR. Hence, the models have attracted attention \cite{Konoplya:2024lch,Liu:2024soc,Liu:2024wal,Malik:2024nhy,Heidari:2024bkm,Wang:2024iwt}.

With the Event Horizon Telescope (EHT) imaging the shadow of the supermassive BH at the center of the M87* galaxy \cite {EventHorizonTelescope:2019pgp}, and the observations of Sgr A* at the center of the Milky Way \cite{EventHorizonTelescope:2022wkp}, BH physics has entered into a new era. These observations not only confirm the existence of BHs, but also offer possibilities for exploring the strong gravitational fields of BHs. The central dark area of BH image is referred to as the BH shadow. By considering that the size and shape of BH shadow depend on the spacetime metric, the study of black hole shadow provides us an important way to extract very useful information about the BH spacetime, and helps us to distinguish various gravitational theories and even to investigate the quantum gravity effects. After the pioneering work by Synge \cite{Synge:1966okc}, several studies on BH shadow have been made in various BH spacetimes, including quantum-corrected ones \cite{Bardeen:1973,Luminet:1979nyg,Johannsen:2013vgc,Amir:2016cen,Abdujabbarov:2016hnw,Shaikh:2019fpu,Gralla:2019xty,Wei:2020ght,Huang:2023yqd,Meng:2023htc,Liu:2020ola,Yang:2022btw,Zhang:2023okw}.

Compared to the static spherically symmetric solution, the exact rotating solution is in general hard to obtain by solving the field equations. An effective method to get a rotating solution is the Newman-Janis algorithm (NJA), which was initially designed to generate the Kerr metric from the Schwarzschild metric in GR \cite{Newman:1965tw, Newman:1965my}.  However, this method encounters some issues. The first one is that complexification of the metric functions is not unique, and there is no rule for choosing the appropriate complexification. The second one is that the generated metric may not be a solution to the same field equations as the seed metric. To avoid introducing complex coordinates during the algorithm process, a modified NJA was proposed in \cite{Azreg-Ainou:2014pra}. The modified NJA makes it easier to obtain physically feasible solutions, and can be directly applied to more complex metric forms, including those obtained within the framework of LQG. Despite these progresses, the modified NJA introduces an undetermined function $H$ to the resulting rotating metric, whose value should be determined by the field equations. Fortunately, the undetermined function $H$ is an overall conformal factor appeared in the rotating metric. Moreover, it turns out that a rotating solution generated by the modified NJA always admits separability of Hamilton-Jacobi equation for null geodesics, irrespective of the ambiguity raised by the undetermined function \cite{Junior:2020lya}. Therefore, neither the causal structure nor the null geodesic depends on the choice of $H$ \cite{Junior:2020lya}. Hence, both the shape and the size of shadow cast by the rotating BH obtained via the modified NJA do not depend on the precise value of $H$. The NJA and its modification have been widely used to generate the corresponding rotating solution from a static seed metric in alternative theories of gravity as well as in quantum gravity \cite{Hansen:2013owa, Shaikh:2019fpu, Brahma:2020eos, Liu:2020ola}.

In this paper, we will use the modified NJA method to generate the corresponding rotating solutions from the two static spherically symmetric quantum-corrected BH solutions proposed in \cite{Zhang:2024khj}, and study the effects of quantum gravity on the shadows cast by the rotating BHs. The structure of our article will be organized as follows. In Sec. \ref{section2}, we recall the two new spherically symmetric quantum-modified BH spacetimes. In Sec. \ref{section3}, we generate the corresponding rotating BHs using the modified NJA, investigate the horizons and the static limit surfaces of these rotating BHs, and analyze how the outer horizons and the static limit surfaces change with the quantum parameters $\zeta$ and the spin parameter $a$. In Sec. \ref{section4}, we discuss the geodesic equations of photons, and study the shadows of the rotating BHs, focusing on the quantum gravity effects. A summary is presented in Sec.\ref{section5}. Throughout this paper, we adopt the geometric unit $G=c=1$, and set the BH mass $M=1$ in numerical calculations.

\section{Static quantum-corrected BHs}\label{section2}
Covariance requires that physical laws have the same form in all coordinate systems. In general relativity, covariance can be satisfied by representing physical quantities as tensors. Similarly, one hopes that covariance can be maintained in effective quantum gravity. Recently, two spherically symmetric BH models addressing the covariance issue have been proposed in effective quantum gravity \cite{Zhang:2024khj}. 

The line element of spherically symmetric spacetime can be expressed in Schwarzschild coordinates as
\begin{equation}
	d s^{2}=-f(r) d t^{2}+ \frac{d r^{2}}{g(r)}+h(r)\left(d \theta^{2}+\sin ^{2} \theta d \phi^{2}\right).\label{mt01}
\end{equation}
The first type of BH model proposed in \cite{Zhang:2024khj}, referred to as BH-I, is described by the line element \eqref{mt01} with
\begin{equation}
\begin{split}
f(r)&=1-\frac{2 M}{r}+\frac{\zeta^{2}}{r^{2}}\left(1-\frac{2 M}{r}\right)^{2}\\
g(r)&=1-\frac{2 M}{r}+\frac{\zeta^{2}}{r^{2}}\left(1-\frac{2 M}{r}\right)^{2},\\
&=\frac{(r-r_{h})(r-r_{0})\left[r^{2}+r_{0}r+(r_{0}^2+\zeta^{2})\right]}{r^{4}},\\
	h(r)&=r^2,
\end{split}
\end{equation}
where
\begin{equation}
\begin{split}
r_{0}&=\frac{-3^{2/3} \zeta^{2}+3^{1 / 3}\left[\zeta^{2}\left(9 M+\sqrt{3}\sqrt{27 M^{2}+\zeta^{2}}\right)\right]^{2 / 3}}{3\left[\zeta^{2}\left(9 M+\sqrt{3} \sqrt{27 M^{2}+\zeta^{2}}\right)\right]^{1 / 3}}, \\
r_{h}&=2 M,
\end{split}
\end{equation}
with $M$ and $\zeta$ being the ADM mass and quantum parameter, respectively. There are two horizons, the outer horizon $r_{h}$ and the interior horizon $r_{0}$ in BH-I. It is evident that the above solution reduces to the Schwarzschild solution as $\zeta\rightarrow0$. The causal structure of BH-I is similar to that of the Reissner-Nordstr\"om BH.

For the second BH model, referred to as BH-II, the line element  is determined by Eq. \eqref{mt01} with  \cite{Zhang:2024khj}
\begin{equation}
\begin{split}
f(r)&=1-\frac{2M}{r},\\  
g(r)&=1-\frac{2 M}{r}+\frac{\zeta^{2}}{r^{2}}\left(1-\frac{2 M}{r}\right)^{2},\\
h(r)&=r^2.
\end{split}
\end{equation}
As opposed to BH-I, $r=r_{0}$ represents a transition surface in BH-II connecting the BH and a white hole, while $r=r_{h}$ denotes the event horizon. Similarly, it can also return to the classical Schwarzschild BH when quantum parameter $\zeta$ is taken as zero. Compared with the classical Schwarzschild BH, BH-II is free of singularity, in which the classical singularity is replaced by a transition surface that connects the BH to a white hole. Meanwhile, in contrast to certain quantum-corrected BHs, the transitional surface entirely exists within the quantum region.

\section{Rotating quantum-corrected BHs}\label{section3}
In this section, we generate two rotating metrics from the seed metrics \eqref{mt01} via the modified NJA proposed in \cite{Azreg-Ainou:2014pra}, and further study some properties of the rotating BHs.

\subsection{Generating the rotating quantum-corrected BHs}
According to the steps of the modified NJA, firstly, one transforms the static solution written in Schwarzschild coordinates to the one in Eddington-Finkelstein coordinates through a coordinate transformation. Secondly, a complex coordinate transformation is performed on the Eddington-Finkelstein coordinate plane, and a physical parameter $a$ is introduced to represent the spin in this process. Then, the modified new procedure is applied to obtain the new rotating metric in Eddington-Finkelstein coordinates. Finally, by a coordinate transform, one rewrites the metric in Boyer-Lindquist coordinates. Some details of the steps are described in \cite{Azreg-Ainou:2014pra}. By the modified NJA with the seed metric \eqref{mt01}, the resulting rotating BH metric in Boyer-Lindquist coordinates reads 
\begin{equation}
\begin{split}
d s^{2}=\frac{H}{\rho^{2}}  {\left[-\left(1-\frac{2 \tilde{M }  r }{\rho^{2}}\right) d t^{2}-\frac{4 a \tilde{M }  r \sin ^{2} \theta}{\rho^{2}} d t d \phi\right.} \\
\left.+\rho^{2} d \theta^{2}+\frac{\rho^{2} d r^{2}}{\Delta}+\frac{\Sigma \sin ^{2} \theta}{\rho^{2}} d \phi^{2}\right],\label{xian}
\end{split}
\end{equation}
where $a = L/M$ is the spin parameter or the angular momentum per unit mass with $M$ and $L$ denoting the ADM mass and angular momentum of the BH respectively, and 
\begin{equation}
\begin{split}
 \rho^{2} &\equiv k(r)+a^{2} \cos ^{2} \theta,\\
 \tilde{M} &\equiv \frac{k(r)-g(r) h(r)}{2 \sqrt{h(r)}} ,\\
  \Delta&\equiv g(r) h(r)+a^{2},\\
 \Sigma &\equiv\left[k(r)+a^{2}\right]^{2}-a^{2} \Delta (r) \sin ^{2} \theta,
\end{split}
\end{equation}
here
\begin{equation}
 k(r)\equiv \sqrt{\frac{g(r)}{f(r)}} h(r).
\end{equation}
Then, we can obtain two rotating quantum-corrected BHs, denoted by RBH-I and RBH-II, by substituting the metric components of the two models BH-I and BH-II into Eq. \eqref{xian} respectively. It should be noted that the function $H(r, a, \theta)$ in Eq. \eqref{xian} is an undetermined function introduced in the modified NJA. In GR, the function $H$ can be determined by certain nonlinear differential equations derived from Einstein's equation as \cite{Azreg-Ainou:2014pra}
\begin{equation} 
H(r,a, \theta)=r^{2}+a^{2} \cos ^{2} \theta.
\end{equation}
In general, it is difficult to find the solution $H$ to the corresponding nonlinear equations derived from the certain field equations. For some theories, such as some effective theories of quantum gravity, which lack underlying field equations, the explicit form of the function $H$ can not be obtained in principle. Nevertheless, two minimal criteria on $H$ in Eq. \eqref{xian} are:
\begin{itemize}
	\item  As $ a \to 0 $, the resulting metric \eqref{xian} should return to the seed metric \eqref{mt01}.
	\item As $\zeta \to 0$, the resulting metric \eqref{xian} should recover the Kerr spacetime, and further, when $a \to 0$, the resulting metric should revert to the Schwarzschild spacetime.
\end{itemize}
Notice that the function $H$ in Eq. \eqref{xian} is an overall conformal factor. Hence, neither the causal structure nor the null geodesic depends on the choice of $H$ \cite{Junior:2020lya}. In this paper, we will work with the function $H$ unspecified, and the results obtained in this paper turn out to be regardless of the choice of $H$.

\subsection{Horizons}
The horizons of the rotating BHs described by Eq. \eqref{xian} are determined by the $g^{rr} =0$, namely,
\begin{equation}
0=\Delta=\frac{(r-r_{h})(r-r_{0})\left[r^{2}+r_{0}r +\left(r_{0}^{2}+\zeta^{2}\right)\right]}{r^2} +a^{2}.\label{delta0}
\end{equation}
The roots of $\Delta$ describe the radii of the horizons. Notice that the two rotating BHs, RBH-I and RBH-II, have the same $\Delta$. Thus, both of them have the same horizons.

\begin{figure*}
	\centering
	\begin{minipage}[t]{0.45\textwidth}
		\centering
		\includegraphics[width=\linewidth, keepaspectratio]{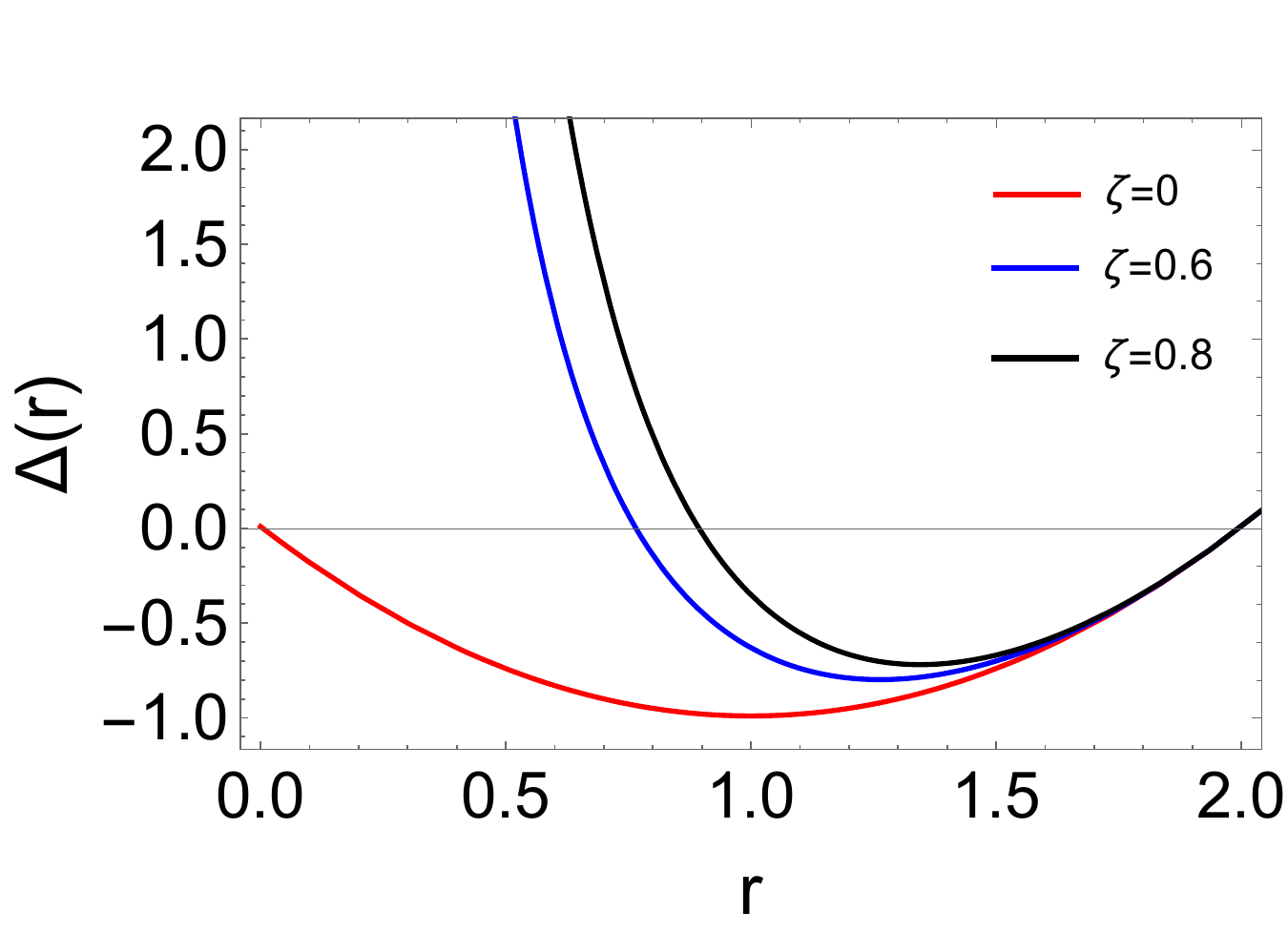}
	\end{minipage}
	\hfill
	\begin{minipage}[t]{0.45\textwidth}
		\centering
		\includegraphics[width=\linewidth, keepaspectratio]{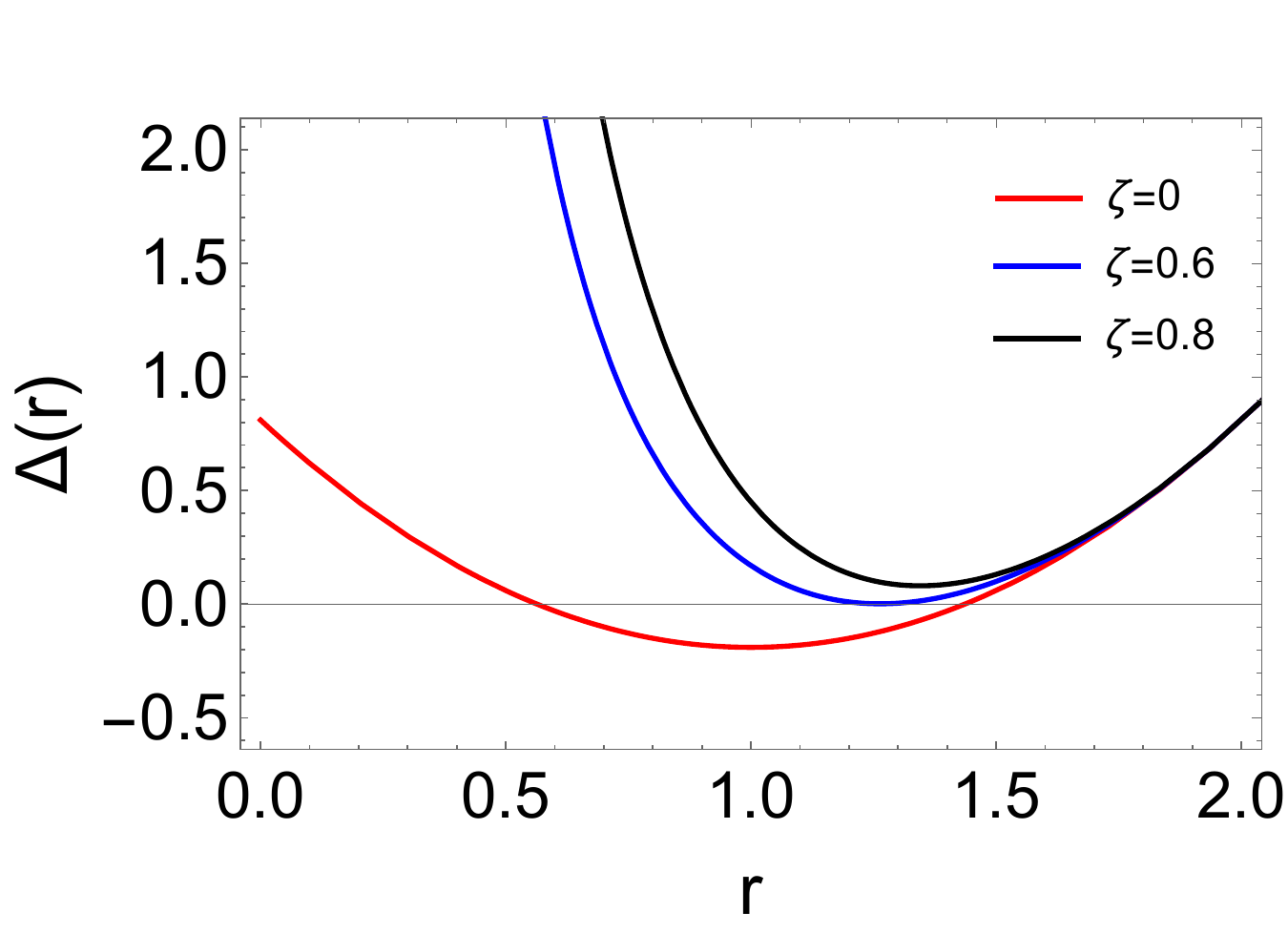}
	\end{minipage}
	\caption{Plots of $\Delta(r)$ with respect to $r$: we fix $a = 0.1$ in the left panel and $a = 0.9$ in the right panel, and vary $\zeta$.}
	\label{fig:EH0}
\end{figure*}

The function $\Delta(r)$ is plotted for different $a$ and $\zeta$ in Fig. \ref{fig:EH0}. The two graphs in Fig.~\ref{fig:EH0} show how the roots of $\Delta$ change with the parameter $\zeta$ in the non-extremal case ($a = 0.1$) and in the near-extremal case ($a = 0.9$). From the left panel in Fig.~\ref{fig:EH0}, we can see that in the non-extremal case, $\Delta$ in Eq. \eqref{delta0} always has two distinct real roots, $r_+$ and $r_-$, corresponding to the outer horizon and the inner horizon respectively, similar to the Kerr BH. As opposed to the outer horizon $r_+$, the inner horizon $r_-$ strongly depends on $\zeta$. The two horizons will get closer and closer as $\zeta$ increases. The right panel in Fig.~\ref{fig:EH0} shows that as the value of $a$ approaches a large value ($a = 0.9$), the parameter $\zeta$ has a significant impact on the roots, and hence on the horizons. In the near-extremal case, the function $\Delta(r)$ has one degenerate root, two real roots, and no real root, depending on the value of $\zeta$. Moreover, in the case of two roots, both $r_+$ and $r_-$ heavily depend on $\zeta$.

\begin{figure}[htbp]
	\centering
	\begin{minipage}[t]{0.45\textwidth}
		\centering
		\includegraphics[width=\textwidth, keepaspectratio]{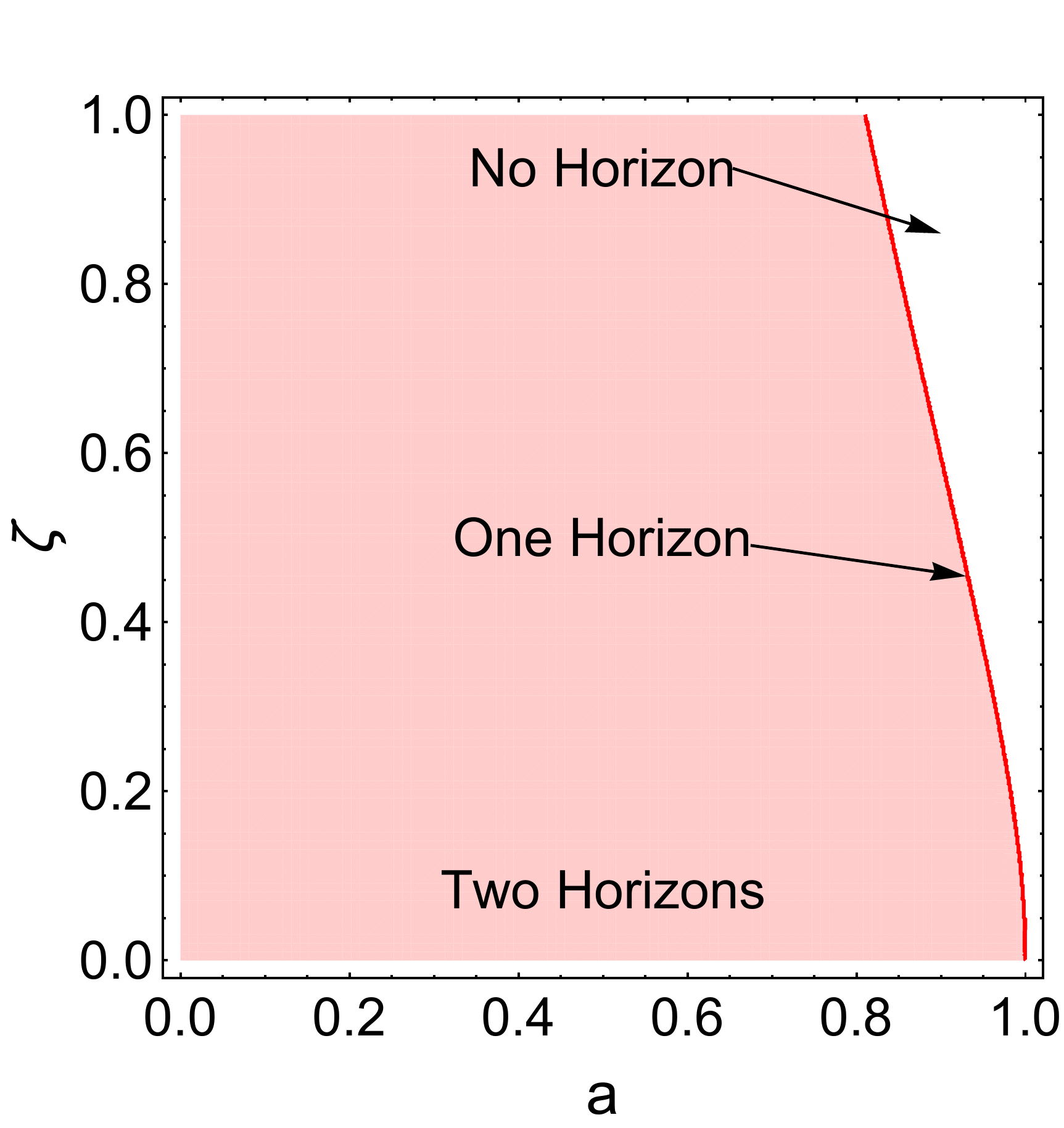}
	\end{minipage}
	\caption{The parameter space $(a,\zeta)$ for two horizons, one degenerate horizon, and no horizon.}
	\label{fig:EH1}
\end{figure}

To clearly understand the variation of the horizons with respect to the parameters, a certain portion of the parameter space of $(a, \zeta)$ for horizons is shown in Fig.~\ref{fig:EH1}. It can be seen that the impact of the quantum parameter $\zeta$ on the horizons is significant only when $a$ is large.

\subsection{The static limit surfaces}

The static limit surfaces, the ergosphere, of the rotating BHs are determined by $g_{tt} = 0$, namely
\begin{equation}
 0=\frac{(r-r_{h})(r-r_{0})\left[r^{2}+r_{0}r +\left(r_{0}^{2}+\zeta^{2}\right)\right]}{r^2}+a^{2}\cos^{2}\theta\equiv\varGamma.\label{equ static}
\end{equation}
The same functions $\varGamma$ for RBH-I and RBH-II lead to the same static limit surfaces. Obviously, the functions $\Delta$ in Eq. \eqref{delta0} and $\varGamma$ in Eq. \eqref{equ static} differ only in the second terms. The function $\Gamma$ can be written in terms of $\Delta$ as
\begin{align}
 \Gamma=\Delta-a^2\sin^2\theta.
\end{align}
Thus, for an arbitrary but fixed $\theta$, the behavior of $\varGamma$ is similar to that of $\Delta$.
\begin{figure*}[htbp]
	\begin{minipage}[t]{0.45\textwidth}
		\centering
		\includegraphics[width=3in, height=3in, keepaspectratio]{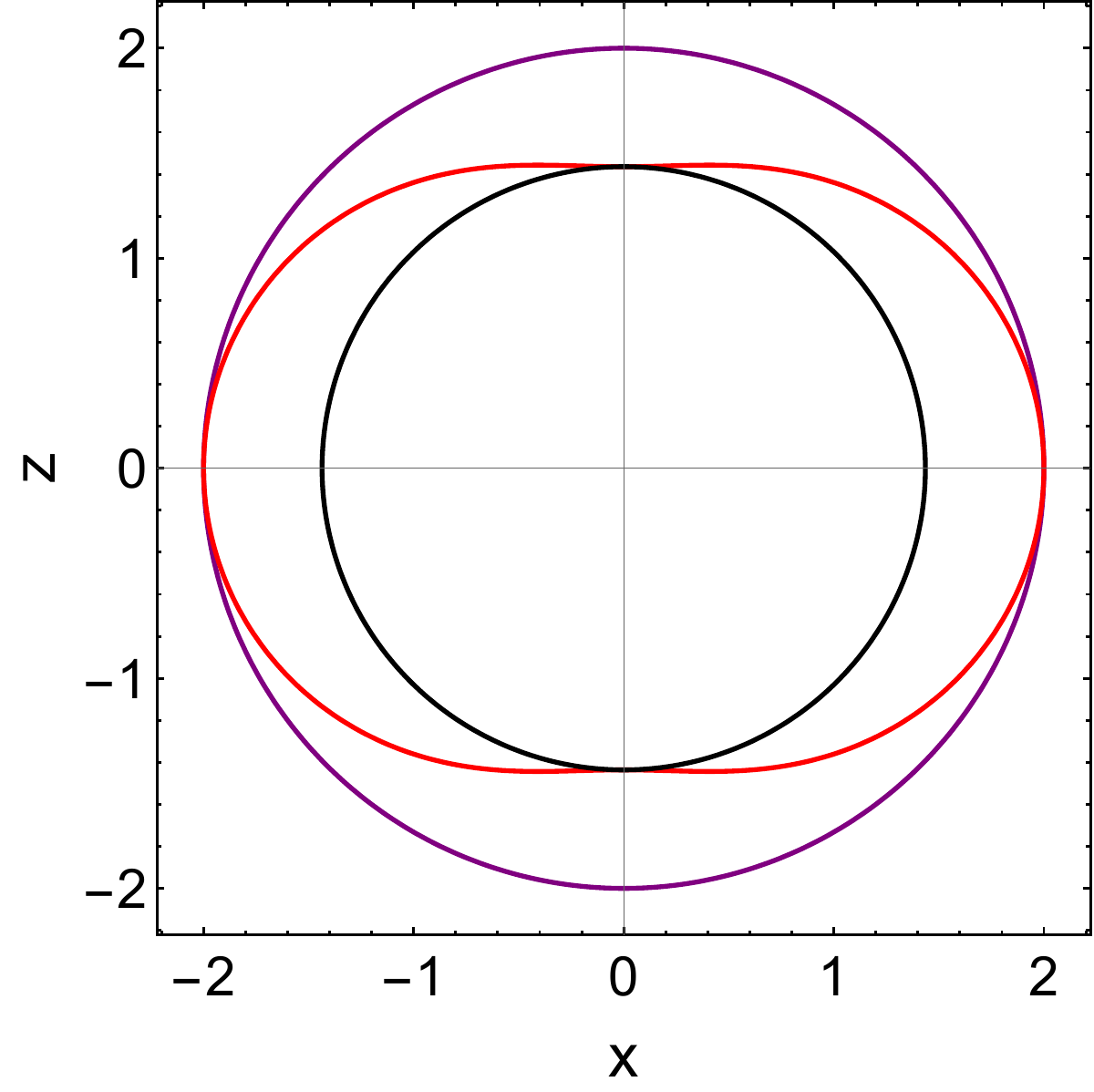}
	\end{minipage}
	\begin{minipage}[t]{0.45\textwidth}
		\centering
		\includegraphics[width=3in, height=3in, keepaspectratio]{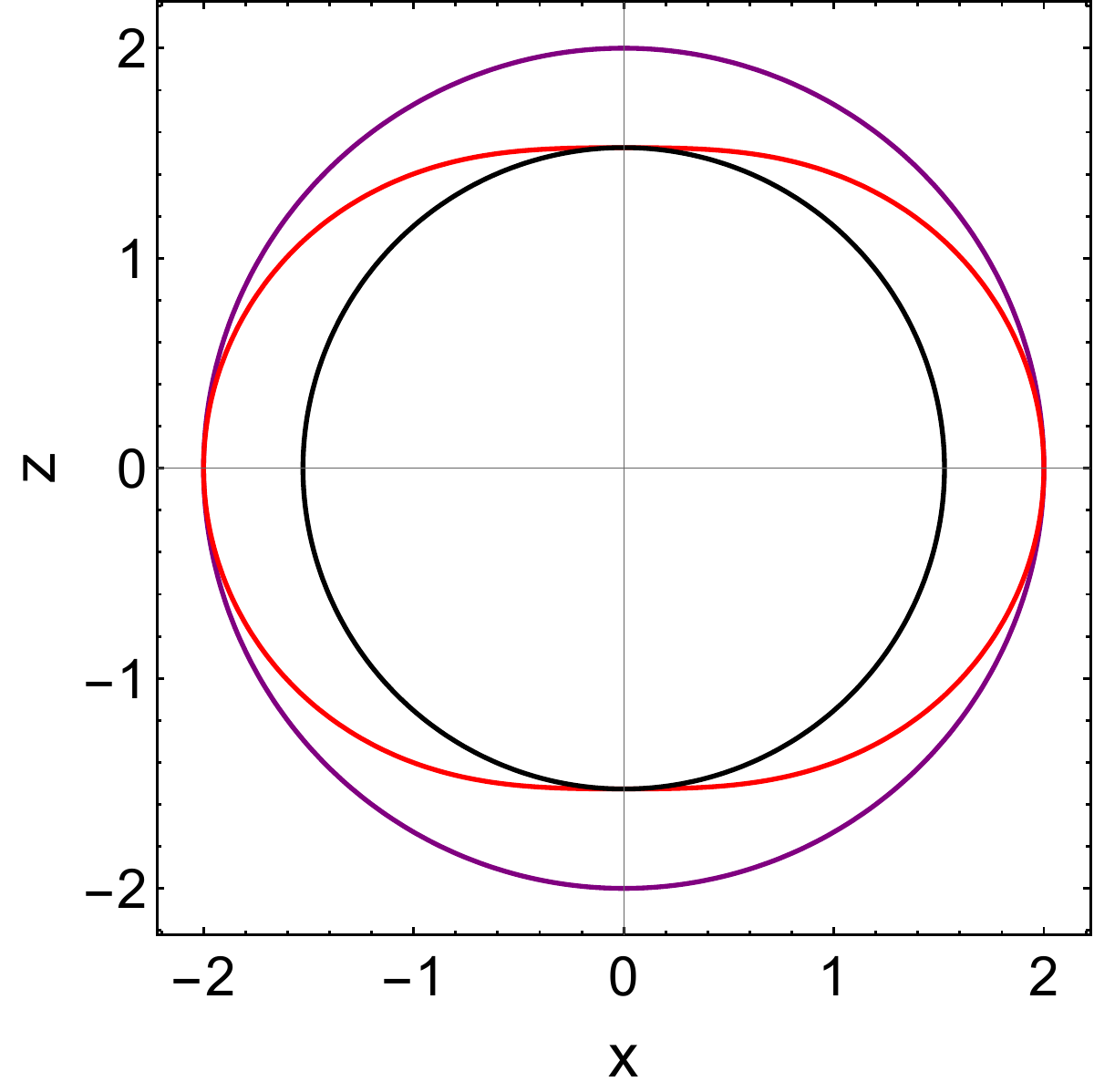}
	\end{minipage}
	\quad
	\begin{minipage}[t]{0.45\textwidth}
		\centering
		\includegraphics[width=3in, height=3in, keepaspectratio]{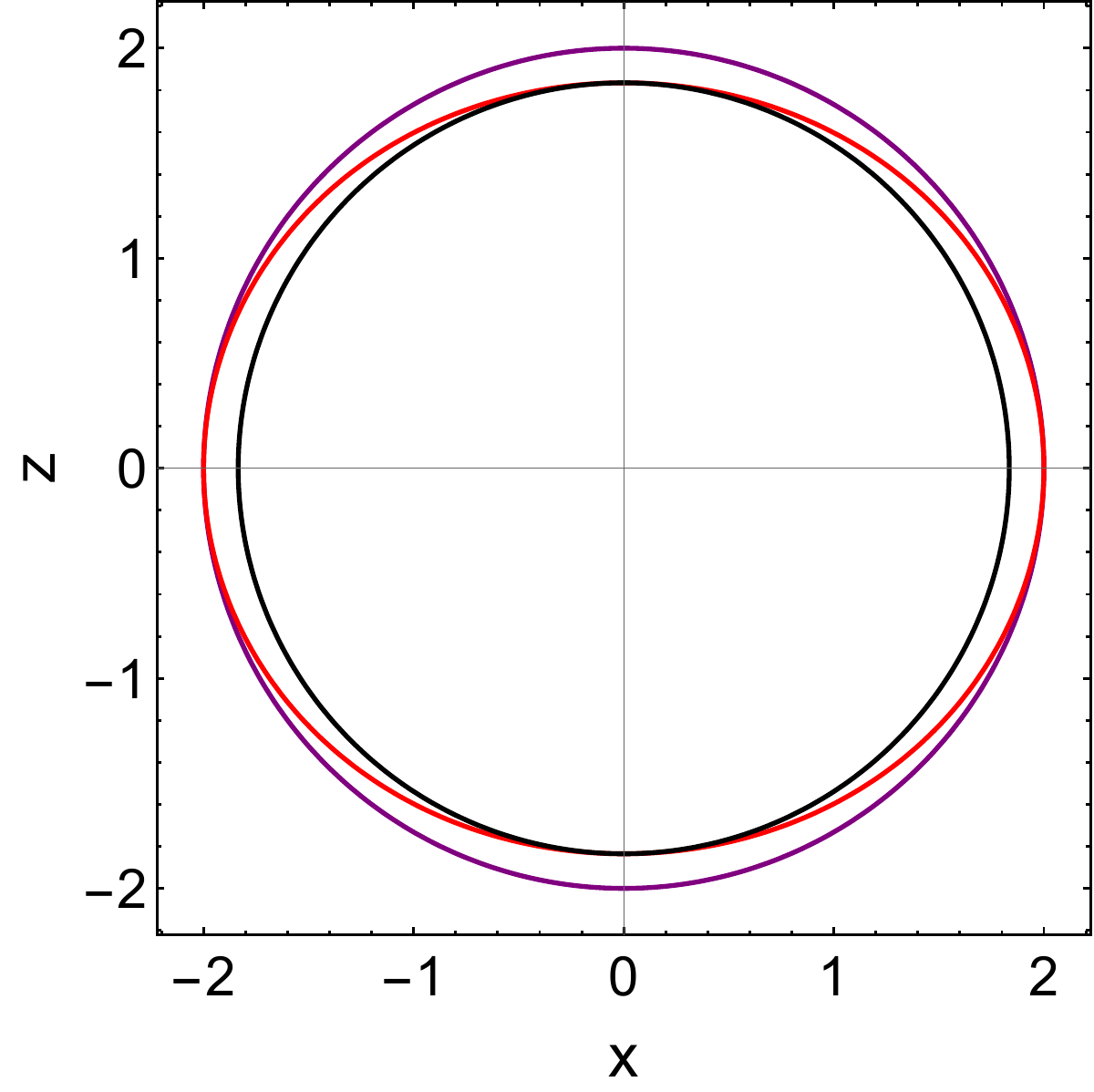}
	\end{minipage}
	\begin{minipage}[t]{0.45\textwidth}
		\centering
		\includegraphics[width=3in, height=3in, keepaspectratio]{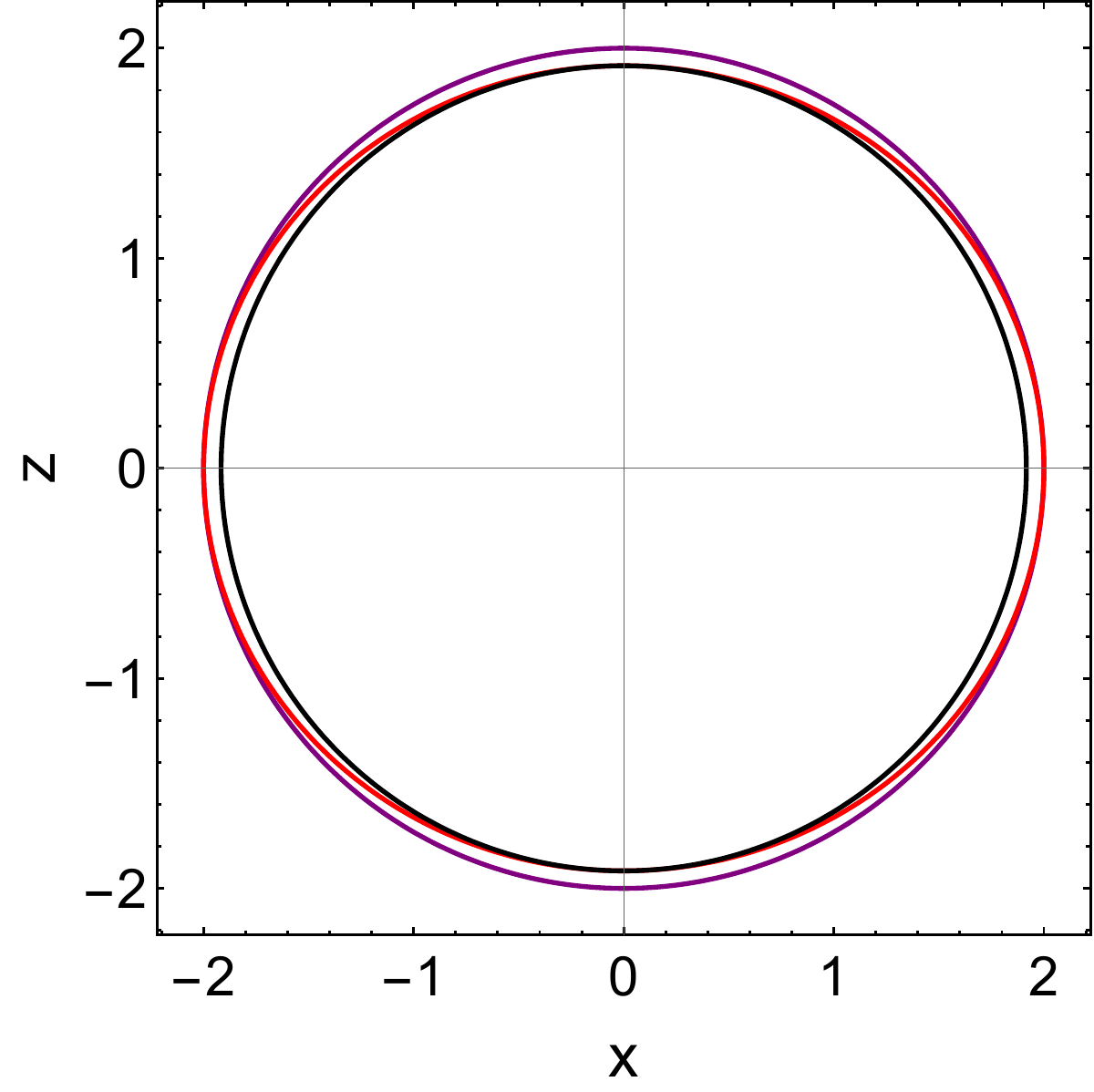}
	\end{minipage}
	\caption{Shapes of the outer horizon (black solid line) and static limit surface (red solid line) of RBH-I in comparison to the outer event horizon in the static case (purple solid line) for $\zeta = 0.01$ and different values of $a$: upper and left panel: $a = 0.90$; upper and right panel: $a = 0.85$; bottom and left panel: $a = 0.55$; and bottom and right panel: $a = 0.40$.}
	\label{fig:erga2}
\end{figure*}

\begin{figure*}[htbp]
	\begin{minipage}[t]{0.45\textwidth}
		\centering
		\includegraphics[width=3in, height=3in, keepaspectratio]{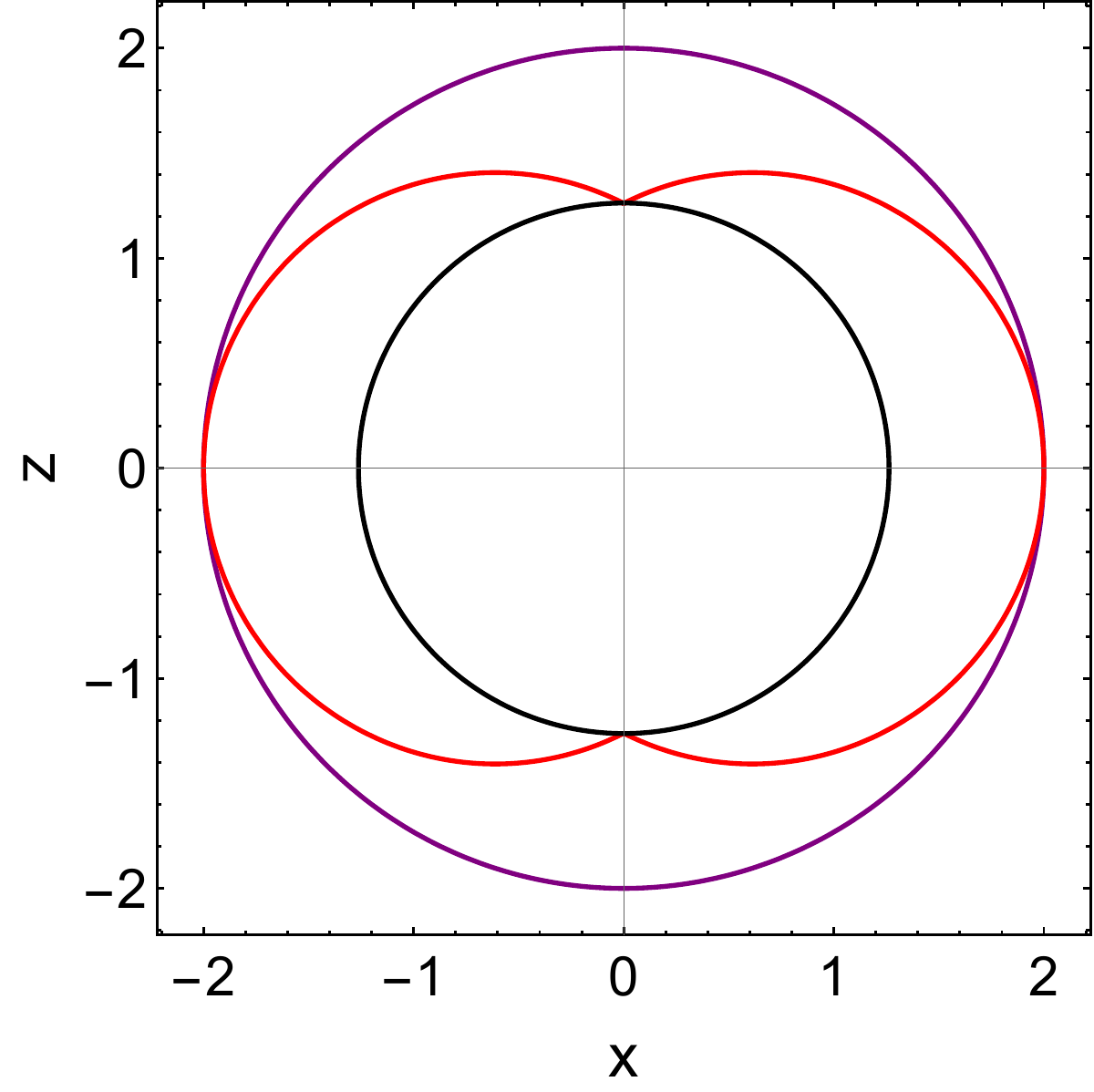}
	\end{minipage}
	\begin{minipage}[t]{0.45\textwidth}
		\centering
		\includegraphics[width=3in, height=3in, keepaspectratio]{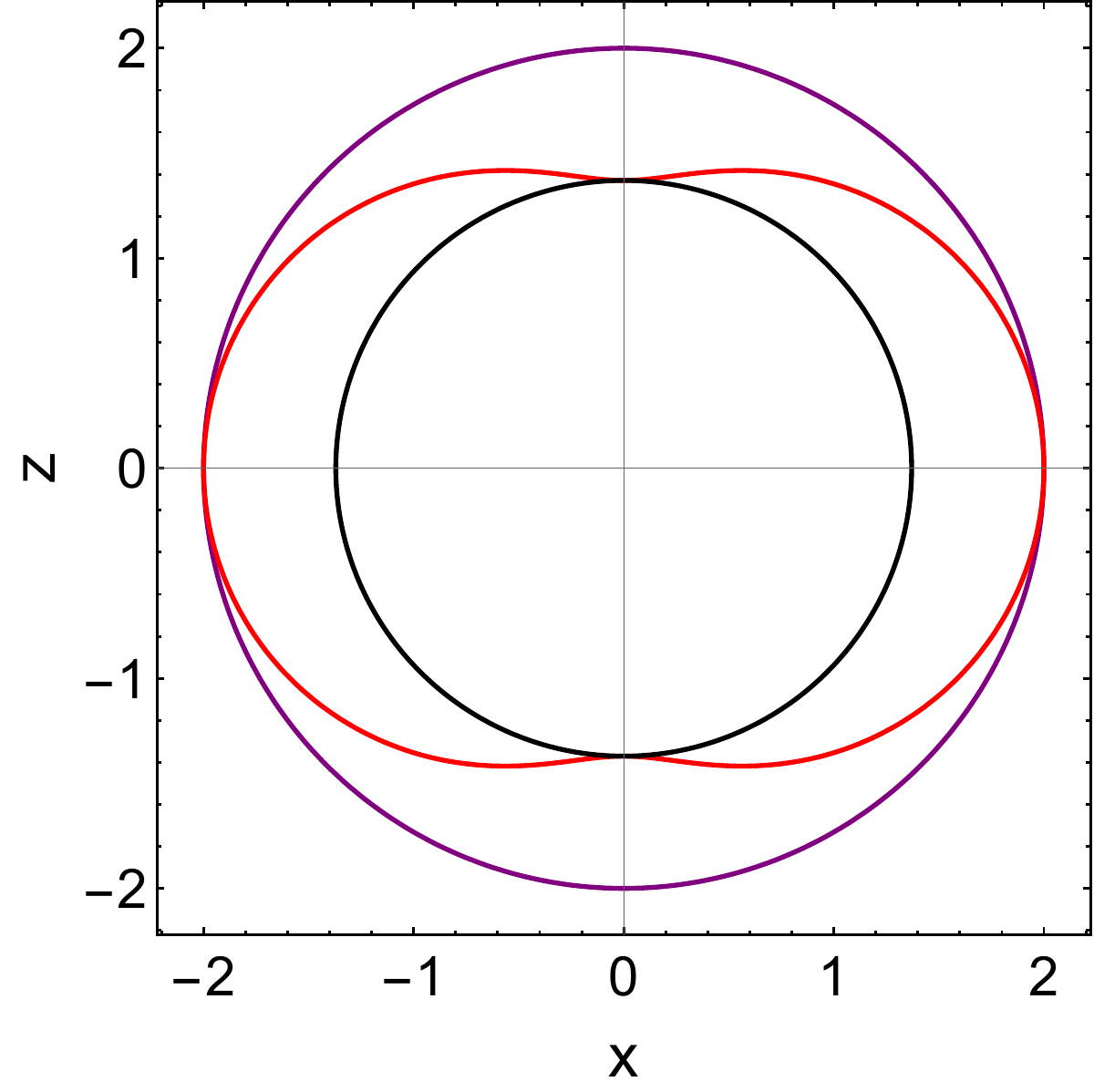}
	\end{minipage}
	\quad
	\begin{minipage}[t]{0.45\textwidth}
		\centering
		\includegraphics[width=3in, height=3in, keepaspectratio]{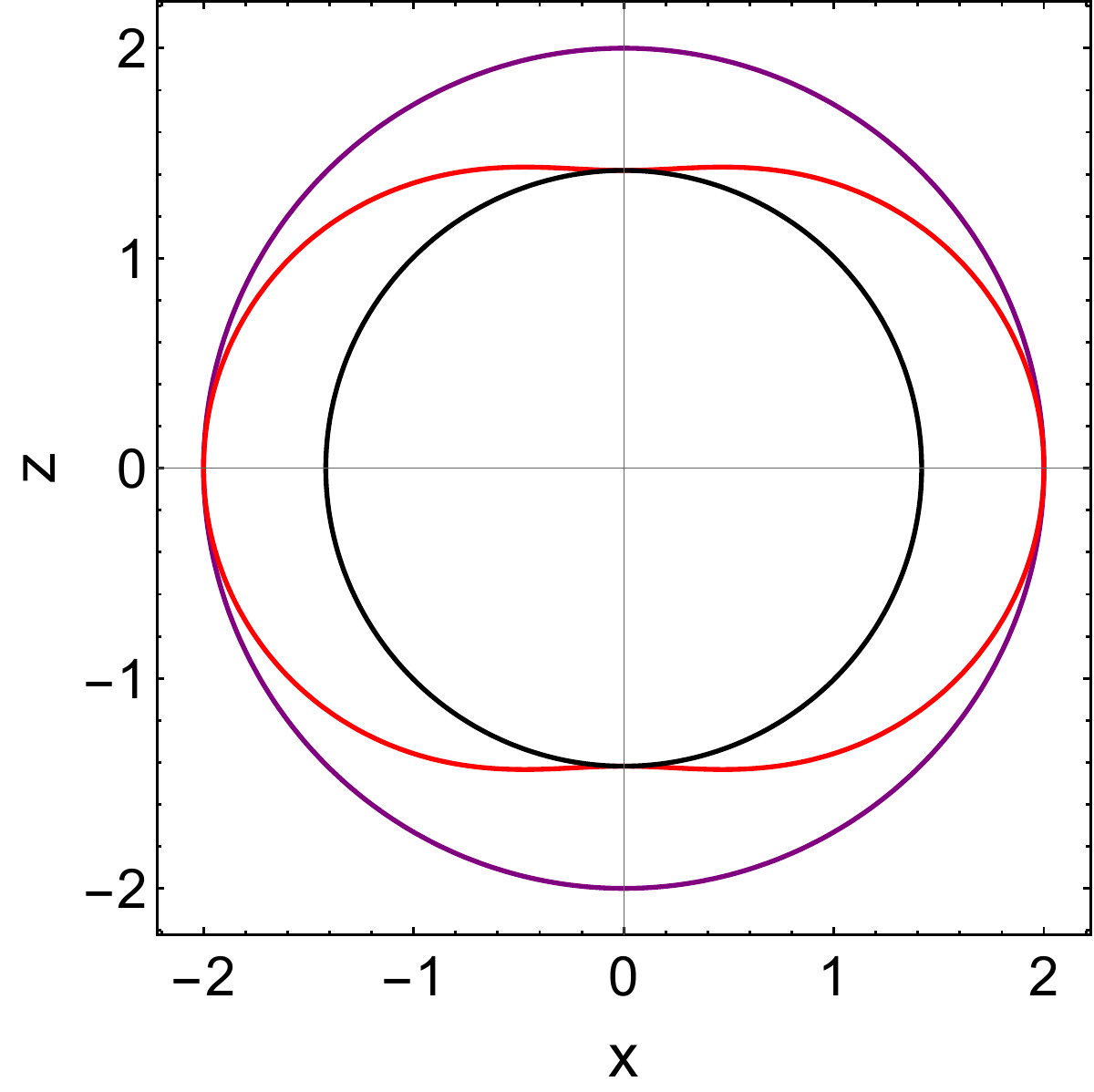}
	\end{minipage}
	\begin{minipage}[t]{0.45\textwidth}
		\centering
		\includegraphics[width=3in, height=3in, keepaspectratio]{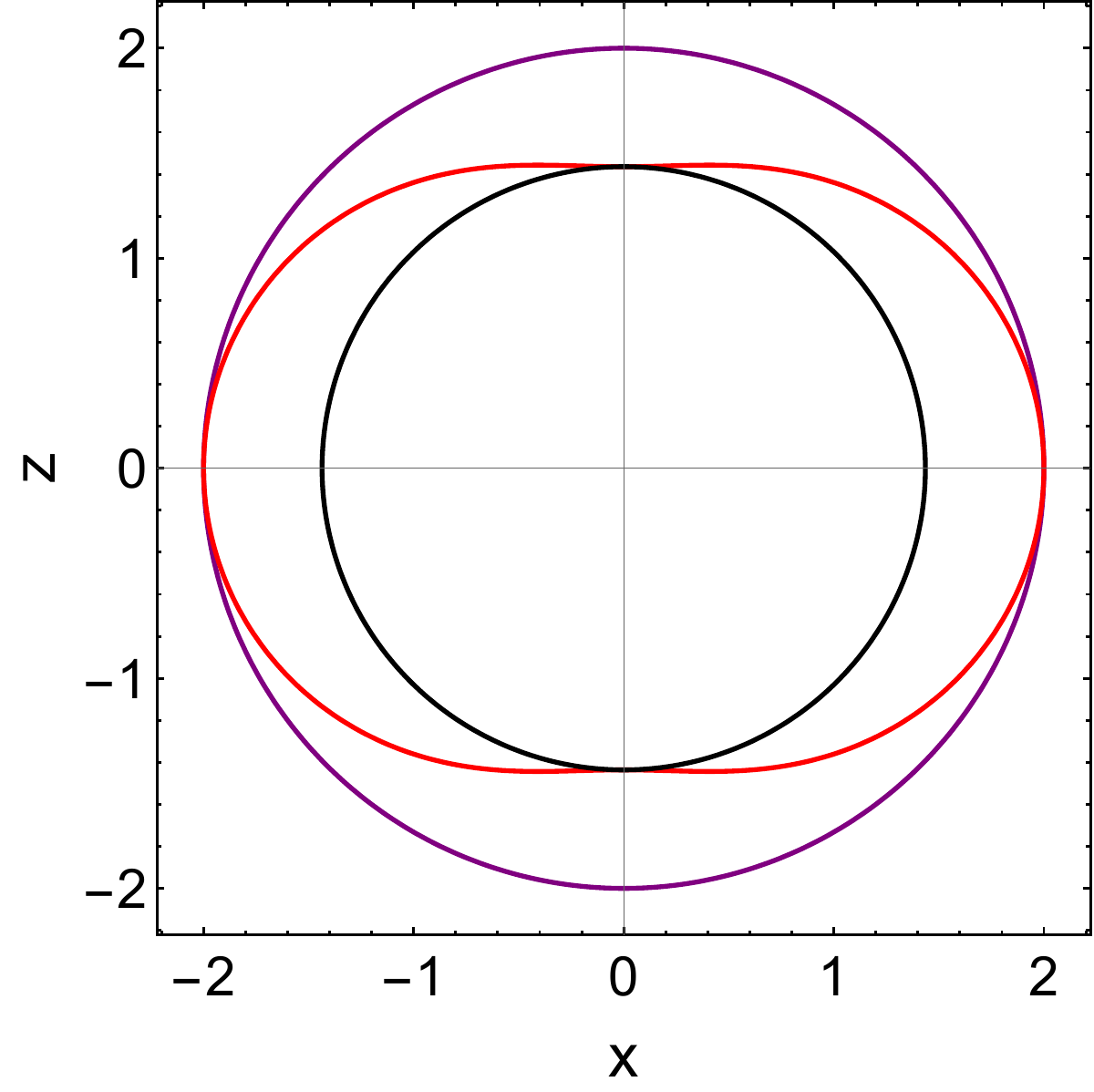}
	\end{minipage}
	\caption{Shapes of the outer horizons (black solid line) and static limit surface (red solid line) of RBH-I in comparison to the outer event horizon in the static case (purple solid line) for $a = 0.9$ and different values of $\zeta$: upper and left panel: $\zeta = 0.59$; upper and right panel: $\zeta = 0.5$; bottom and left panel: $\zeta = 0.3$; and bottom and right panel: $\zeta = 0.001$.}
	\label{fig:erga3}
\end{figure*}
Considering Cartesian coordinates $x=r\sin\theta\cos\phi$, $y=r\sin\theta\sin\phi$, $z=r\cos\theta$ and the axisymmetry of the rotating BHs, in Figs.~\ref{fig:erga2} and \ref{fig:erga3}, we display the outer horizon $r_+$ and the outer static limit surface of RBH-I (the situation of RBH-II is the same) in the $x-z$ plane for different parameters $a$ and $\zeta$. In these figures, the purple circle corresponds to the outer horizon $r_h$ in the static case with $a = 0$. From Fig.~\ref{fig:erga2}, for a fixed $\zeta=0.01$, it can be seen that the outer horizon $r_+$ and the outer static limit surface monotonically and rapidly increase as $a$ decreases, and gradually approach the outer horizon $r_h$ in the static case. From Fig. \ref{fig:erga3}, for a fixed $a=0.9$, we observe that both the outer horizon $r_+$ and the outer static limit surface slightly increase as $\zeta$ decreases. Thus, among $a$ and $\zeta$, the spin parameter $a$ has a significant influence on the static limit surface.

\section{Null geodesic around the rotating quantum-corrected BHs and BH shadows}\label{section4}
In this section, we aim to further analyze the relationship between these rotating quantum-corrected BHs generated in the previous section and the Kerr BH through the study of BH shadows, and examine the influence of $\zeta$ on the BH shadow. The contour of a BH shadow is defined by the photon orbits around it. Therefore, in order to obtain the rotating BH shadow, one first needs to study the equation of motion for photons in the background spacetime.  
\subsection{Null geodesic equations}
In the spacetime \eqref{xian}, there exist two Killing vector fields, the timelike Killing vector field $\left(\partial/\partial t\right)^{a}$ and the axial Killing vector field $\left(\partial/\partial \phi\right)^{a}$, leading to two conserved quantities for null geodesics, the energy $E$ and the angular momentum $L$ \cite{Wald:1984bk}
\begin{equation}
	\begin{split}
		&E:=-g_{a b}\left(\frac{\partial}{\partial t}\right)^{a}\left(\frac{\partial}{\partial \tau }\right)^{b}=-g_{t t} \frac{\mathrm{d} t}{\mathrm{~d} \tau }-g_{t\phi}\frac{\mathrm{d} \phi}{\mathrm{~d} \tau },\\
		&L:=g_{a b}\left(\frac{\partial}{\partial \phi}\right)^{a}\left(\frac{\partial}{\partial \tau }\right)^{b}=g_{\phi \phi} \frac{\mathrm{d} \phi}{\mathrm{d} \tau }+g_{t\phi}\frac{\mathrm{d} t}{\mathrm{~d} \tau },\label{EL}
	\end{split}
\end{equation}
where $\tau$ represents the affine parameter. The geodesics and orbits in the Kerr spacetime have been first studied by Carter using the Hamilton-Jacobi equation in \cite{Carter:1968rr}. The separability of the Hamilton-Jacobi equation for geodesics was used to obtain an additional constant of motion, the so-called Carter constant \cite{Carter:1968rr}. Indeed, it turns out that the Carter constant can be obtained by using the Killing tensor $K_{ab}$ in the Kerr spacetime \cite{Wald:1984bk}. The Hamilton-Jacobi equation takes the form
\begin{equation}
	\frac{\partial S}{\partial \tau }=-\frac{1}{2} g^{\mu \nu} \frac{\partial S}{\partial x^{\mu}} \frac{\partial S}{\partial x^{\nu}},\label{s31}
\end{equation}
where $S$ denotes the Jacobi action. It turns out that for null geodesics, the rotating spacetime generated via the modified NJA always admits separability in the Hamitlon-Jacobi equation \cite{Junior:2020lya}. Thus, $S$ can be expressed in a separable form as
\begin{equation}
	S=\frac{1}{2} m^{2} \tau -E t+L \phi+S_{r}(r)+S_{\theta}(\theta),  \label{s32}            
\end{equation}
where $m$ represents the mass of the test particle, and it takes the value of 0 for the photon. Substituting Eq.~\eqref{s32} into Eq. \eqref{s31} and separating out the two parts of $r$ and $\theta$ being equal to the Carter constant, we obtain the null geodesics equations for $r$ and $\theta$ determined by the metric \eqref{xian} as
\begin{align}
	\left(H\frac{d r}{d \tau }\right)^{2}&=\left[\left(k+a^{2}\right)E-a L\right]^{2}-\Delta\left[C+(L-a  E)^{2}\right]\notag\\
	&\equiv R(r), \label{likenull3}\\
	\left(H \frac{d \theta}{d \tau }\right)^{2}&=C+a^{2} E^{2} \cos^{2} \theta-L^{2}\cot^{2}\theta\notag\\
	&\equiv\Theta(\theta), \label{likenull4}
\end{align}
where $C$ is the Carter constant, and we have used $\frac{\partial S}{\partial x^{\mu}} = p_{\mu}$ and $p^{\mu} = g^{\mu \nu} p_{\nu} = \dot{x}^{\mu} = \frac{d x^{\mu}}{d \tau}$ with \(p^{\mu}\) being the  momentum.. Solving Eq.~\eqref{EL} with the metric \eqref{xian} yields
\begin{align}
H \frac{d t}{d \tau} &=  a \Big(L - a E\,\sin^2\theta\Big) \Bigg. 
\Bigg. + \frac{k(r) + a^2}{\Delta(r)} \Big[\Big(k(r) + a^2\Big) E - a L\Big] ,\\
H \frac{d \phi}{d \tau }&= {\left ( \frac{L}{\sin ^{2} \theta}-a E\right.} \left.+ \frac{a}{\Delta}\left[\left(k(r)+a^{2}\right) E-a L\right]  \right ) .\label{likenull2}
\end{align}
The null geodesic equations \eqref{likenull3}-\eqref{likenull2} describe how the photons move around the rotating BHs. It is convenient to introduce two impact parameters \cite{Bardeen:1973}
\begin{equation}
\xi=\frac{L}{E}, \quad \eta=\frac{C}{E^{2}}.\label{eq:twoimpactpara}
\end{equation}
The trajectory of a photon is completely determined by its impact parameters $\xi$ and $\eta$.

\subsection{Shadows of rotating quantum-corrected BHs}

To determine the shadow cast by a BH, one needs to specify the ingredients: light source and observer. In this paper, for simplicity as done in \cite{Bardeen:1973}, we consider a light source behind the BH whose angular size is large compared to that of the BH, and a distant observer located in front of the BH. It turns out that, for photons from the light source with varying impact parameters, there are three possible types of orbits near a rotating BH: scatter orbits,  capture orbits and unstable orbits \cite{Bardeen:1973,Amir:2016cen}. The scatter orbits allow photons to scatter away from the BH to infinity. The capture orbits cause photons to fall directly into the BH. The unstable orbits lie between the former two types. For the photons on unstable orbits, small perturbations will cause the photons to escape to infinity or to fall into the BH, forming a boundary that separates the bright and black regions as seen by the observer. The boundary is often referred to as the contour of the BH shadow. The unstable orbits of photons are determined by \cite{Hioki:2009na}
\begin{equation}
\begin{split}
R\left(r_{s}\right) =
\left.\frac{d R(r)}{d r}\right|_{r=r_{s}}  =0    , 
\end{split}\label{R}
\end{equation}
\begin{equation}
\begin{split}
\left.\frac{d^{2} R(r)}{d r^{2}}\right|_{r=r_{s}}  >0 ,
\end{split}\label{Rr}
\end{equation}
where $r_s$ denotes the radius of the photon sphere. Rewriting $R$ in Eq. \eqref{likenull3} in terms of $\xi$ and $\eta$ in Eq. \eqref{eq:twoimpactpara} and then solving Eq.~\eqref{R} for $\xi$ and $\eta$ parametrized in terms of $r_s$, we have  
\begin{equation}
\begin{split}
\xi=\frac{a^{2}+k(r_s)-\frac{2 \Delta(r_s) k^{\prime}(r_s)}{\Delta^{\prime}(r_s)}}{a}, 
\end{split}\label{xi1}
\end{equation}

\begin{widetext}
\begin{equation}
\begin{split}
\eta=\frac{-4 \Delta(r_s)^{2} {k}^{\prime}(r_s)^{2}-{k}(r_s)^{2} \Delta^{\prime}(r_s)^{2}+4 \Delta(r_s) {k}^{\prime}(r_s)\left({a}^{2} {k}^{\prime}(r_s)+{k}(r_s) \Delta^{\prime}(r_s)\right)}{{a}^{2} \Delta^{\prime}(r_s)^{2}} .\\ 
\end{split}\label{eta1}
\end{equation}
\end{widetext}
Consider an observer located at $r=r_*\rightarrow\infty$ and seen with a sight angle $\theta_*$ between the line of sight and the rotation axis of the BH. Then, the contour of the BH shadow seen by the observer can be determined by the celestial coordinates
\begin{align}
	X & =\lim _{r{\ast} \rightarrow \infty}\left(-\left.r_{\ast}^{2} \sin \theta_{\ast} \frac{d \phi}{d r}\right|_{\theta \rightarrow \theta_{\ast}}\right)=-\xi \csc \theta_{\ast}, \label{eq:X}\\
	Y & =\lim _{r_{\ast} \rightarrow \infty}\left(\left.r_{\ast}^{2} \frac{d \theta}{d r}\right|_{\theta \rightarrow \theta_{\ast}}\right)  = \pm \sqrt{\eta+a^{2} \cos ^{2} \theta_{\ast}-\xi^{2} \cot ^{2} \theta_{\ast}}, \label{eq:Y}
\end{align}
where in the second step of Eq. \eqref{eq:X}, the null geodesic equations \eqref{likenull2} and \eqref{likenull3} were used, and Eqs. \eqref{likenull4} and \eqref{likenull3} were inserted in \eqref{eq:Y}. It is worth noting that the undetermined factor $H$ commonly appeared in the null geodesic equations has been canceled in the resulting expressions of $X$ and $Y$. Thus, the rotating BH shadows are regardless of the choice of $H$, as pointed out in \cite{Junior:2020lya}. Considering the observer on the equatorial plane of the rotating BHs with $\theta_{\ast} = \frac{\pi}{2}$, we have
\begin{equation}
\begin{split}
X =-\xi , \quad
Y = \pm \sqrt{\eta} .\label{XY}
\end{split}
\end{equation}
In what follows, we will focus on this case.

\begin{figure}[htbp]
	\centering
	\begin{minipage}[t]{0.45\textwidth}
		\centering
		\includegraphics[width=\textwidth, keepaspectratio]{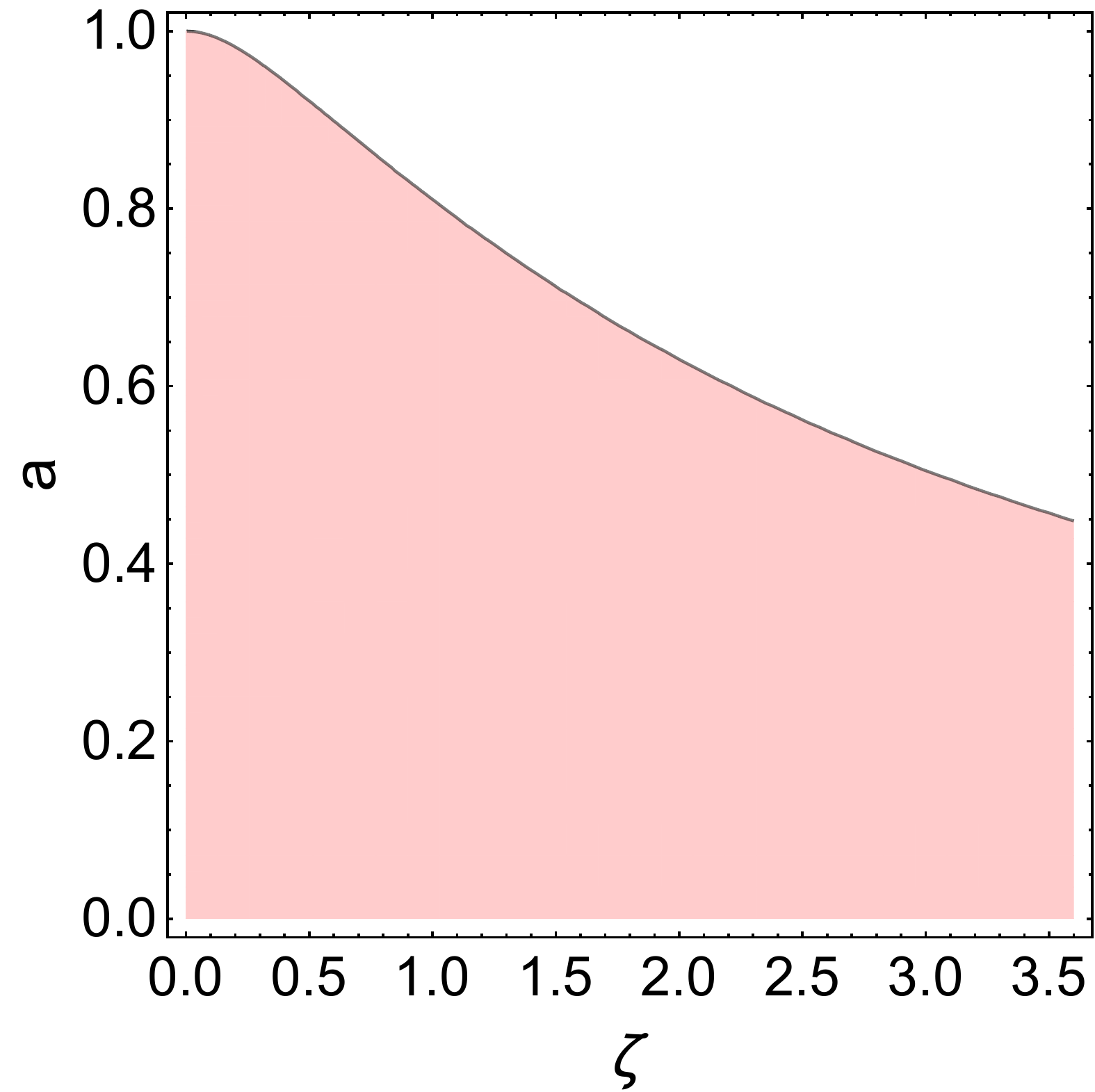}
	\end{minipage}
	\caption{Parameter space of $(\zeta,a)$ allowing the existence of the shadows cast by RBH-I and RBH-II. }
	\label{fig:a_zeta1}
\end{figure}

It is not difficult to see that the obtained celestial coordinates are constrained by \(a\) and \(\zeta\). Therefore, in order to get accurate BH shadows, one needs first to constrain the parameters \(a\) and \(\zeta\). The same parameter space of $(\zeta,a)$ allowing the exsitence of shadows cast by RBH-I and RBH-II is plotted in Fig. \ref{fig:a_zeta1}.

\begin{figure*}[htbp]
	\begin{minipage}[t]{0.45\textwidth}
		\centering
		\includegraphics[width=3in, height=3in, keepaspectratio]{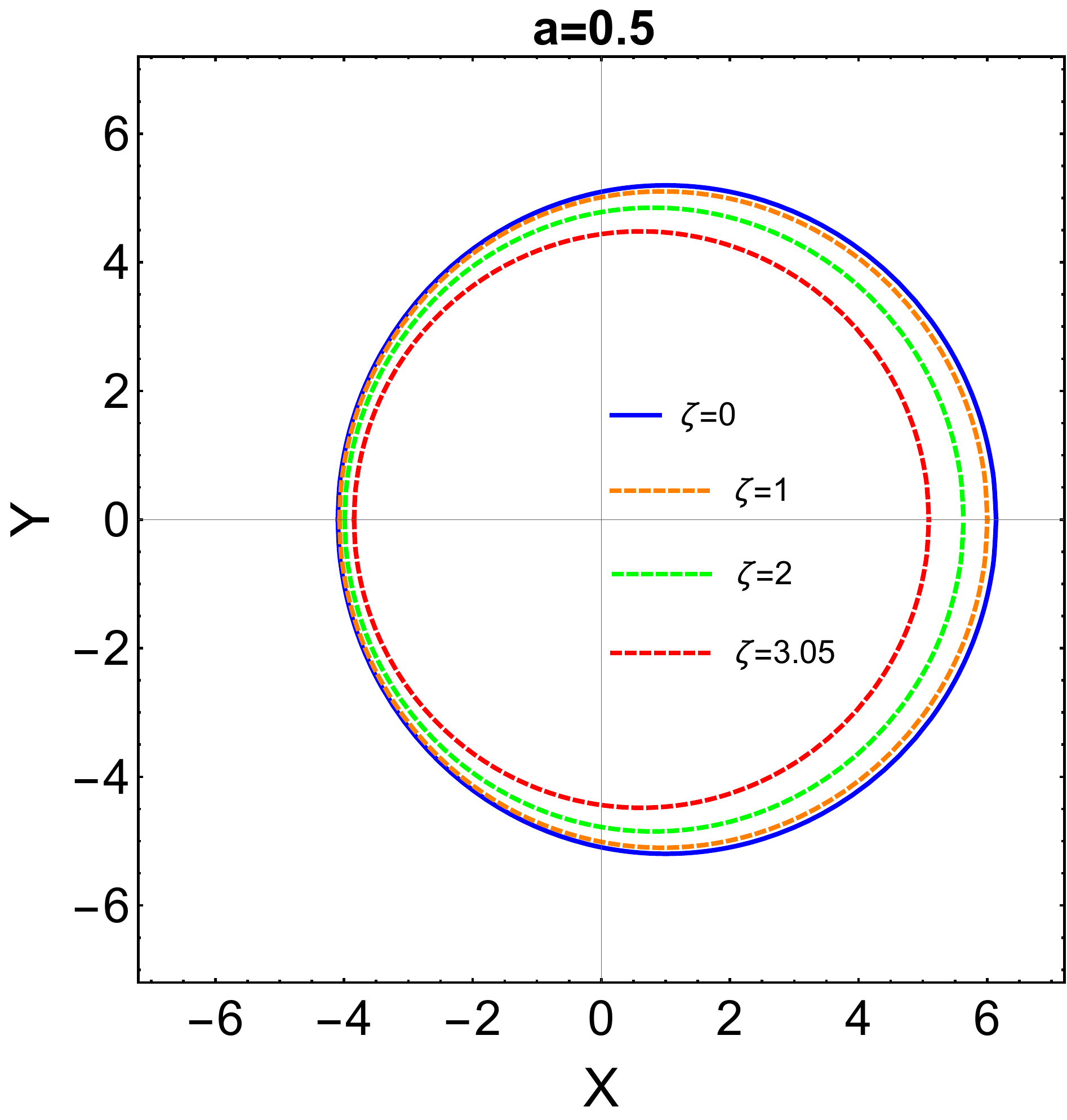}
	\end{minipage}
	\begin{minipage}[t]{0.45\textwidth}
		\centering
		\includegraphics[width=3in, height=3in, keepaspectratio]{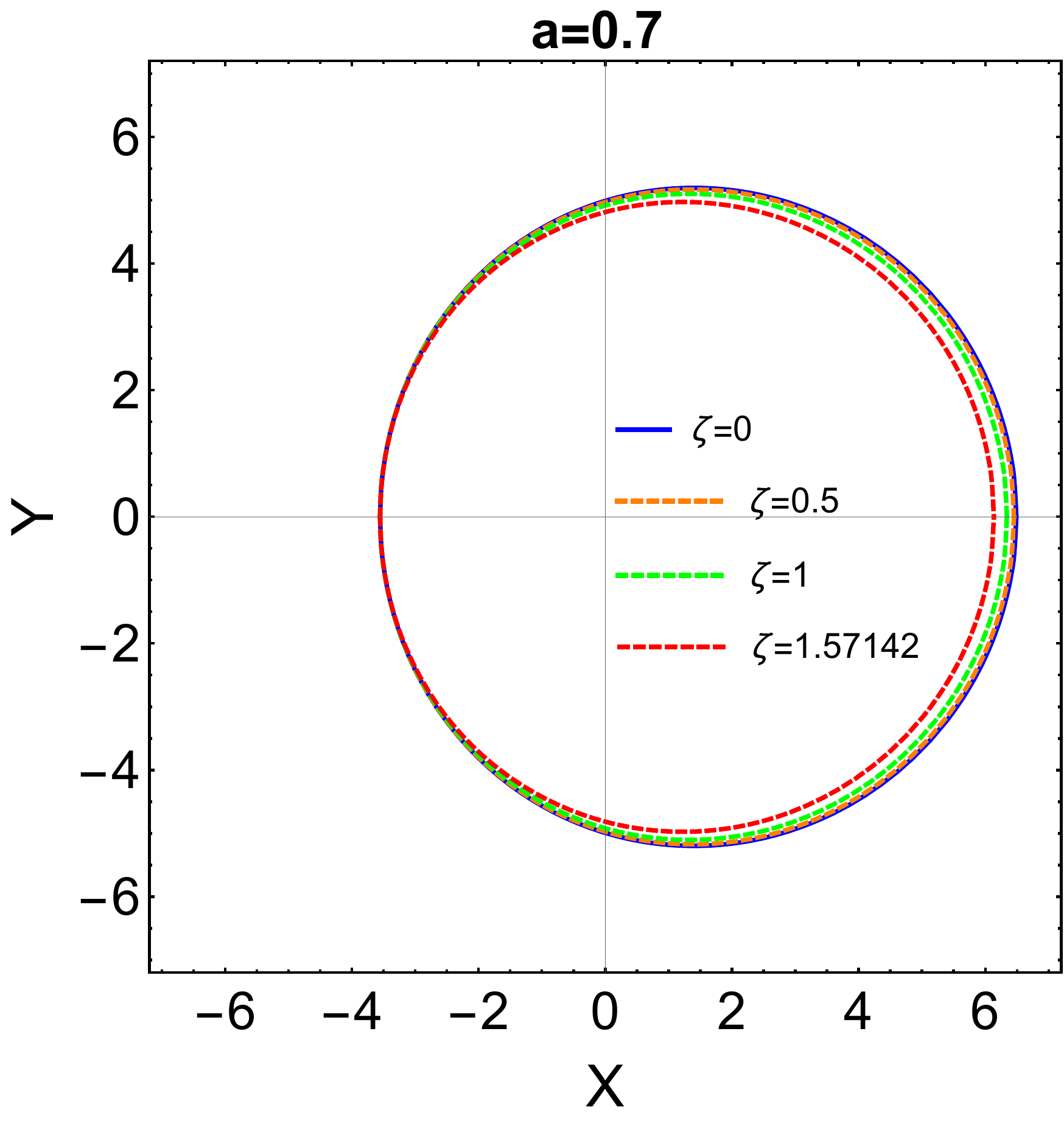}
	\end{minipage}
	\quad
	\begin{minipage}[t]{0.45\textwidth}
		\centering
		\includegraphics[width=3in, height=3in, keepaspectratio]{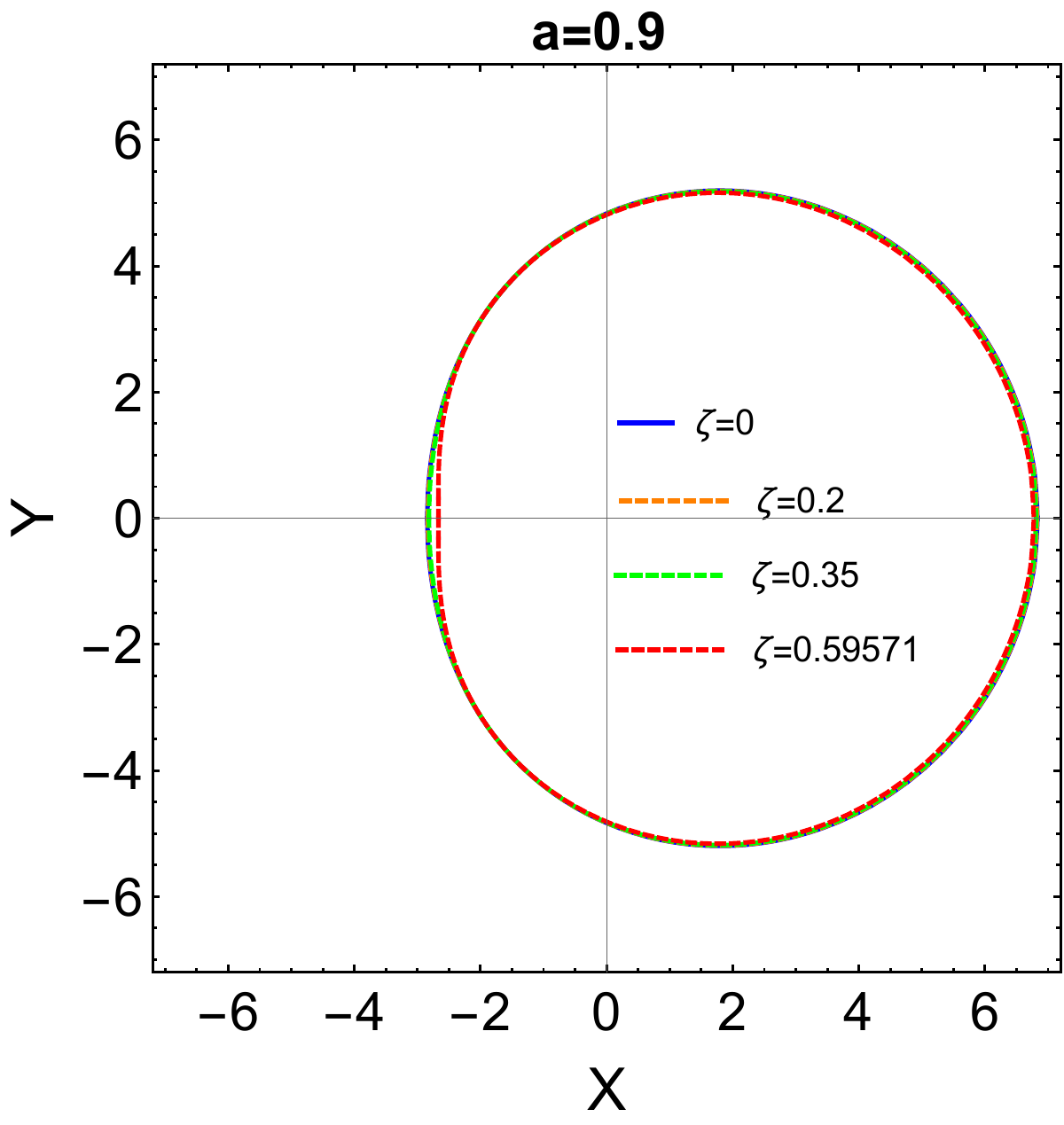}
	\end{minipage}
	\begin{minipage}[t]{0.45\textwidth}
		\centering
		\includegraphics[width=3in, height=3in, keepaspectratio]{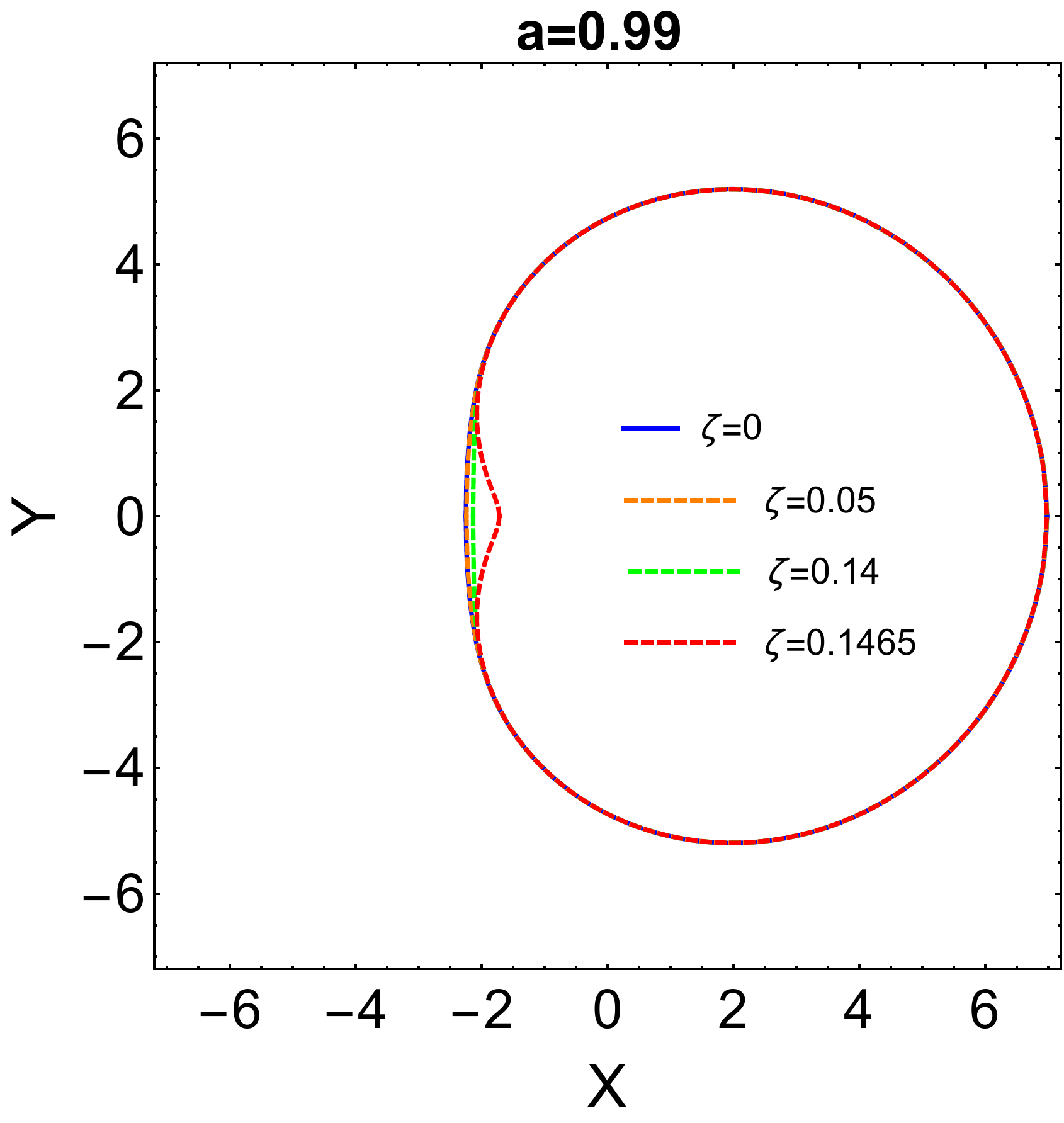}
	\end{minipage}
	\caption{Shadows cast by RBH-I for different parameters $a$ and $\zeta$.}
	\label{fig:rbh1}
\end{figure*}

\begin{figure*}[htbp]
	\begin{minipage}[t]{0.45\textwidth}
		\centering
		\includegraphics[width=3in, height=3in, keepaspectratio]{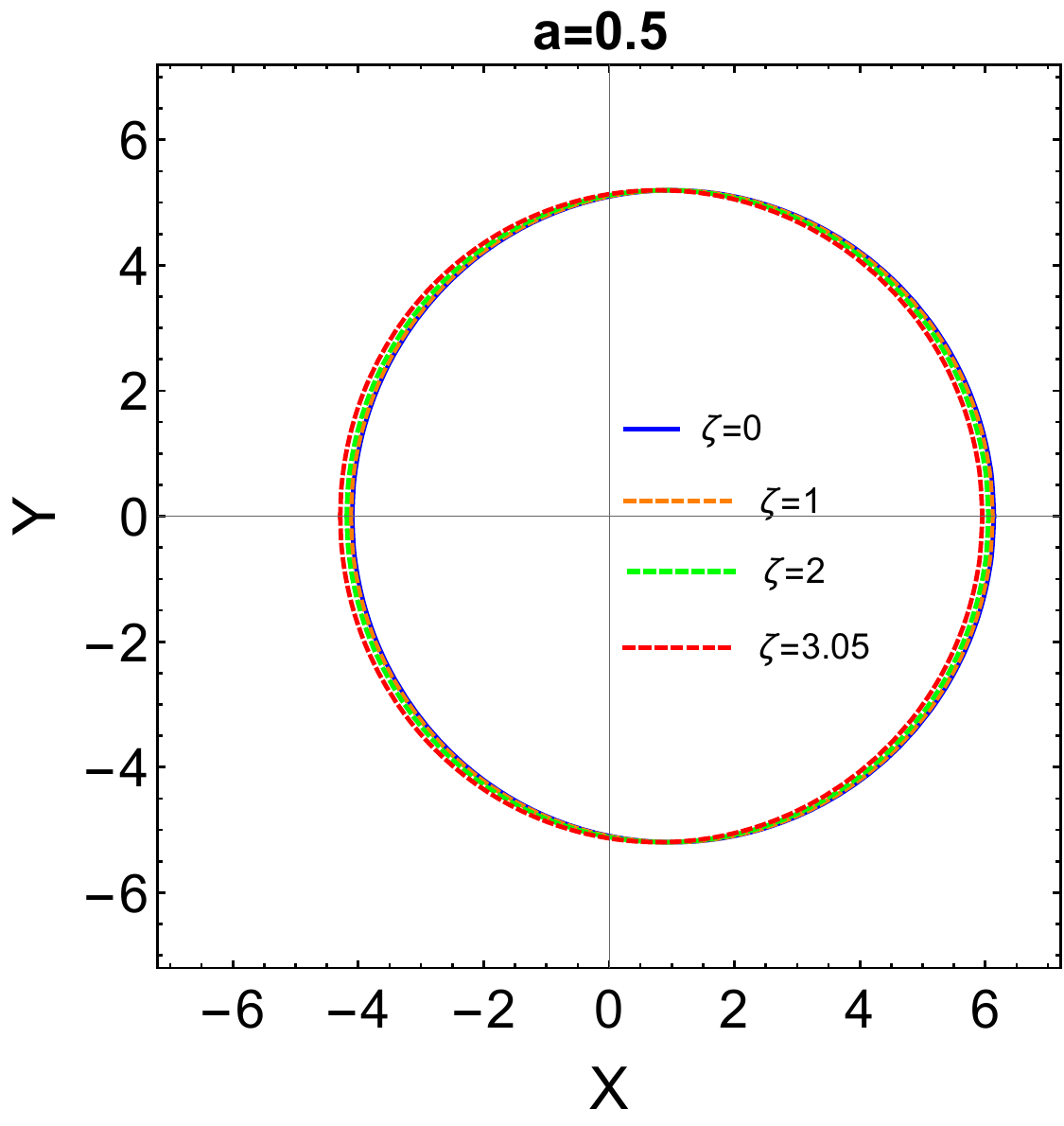}
	\end{minipage}
	\begin{minipage}[t]{0.45\textwidth}
		\centering
		\includegraphics[width=3in, height=3in, keepaspectratio]{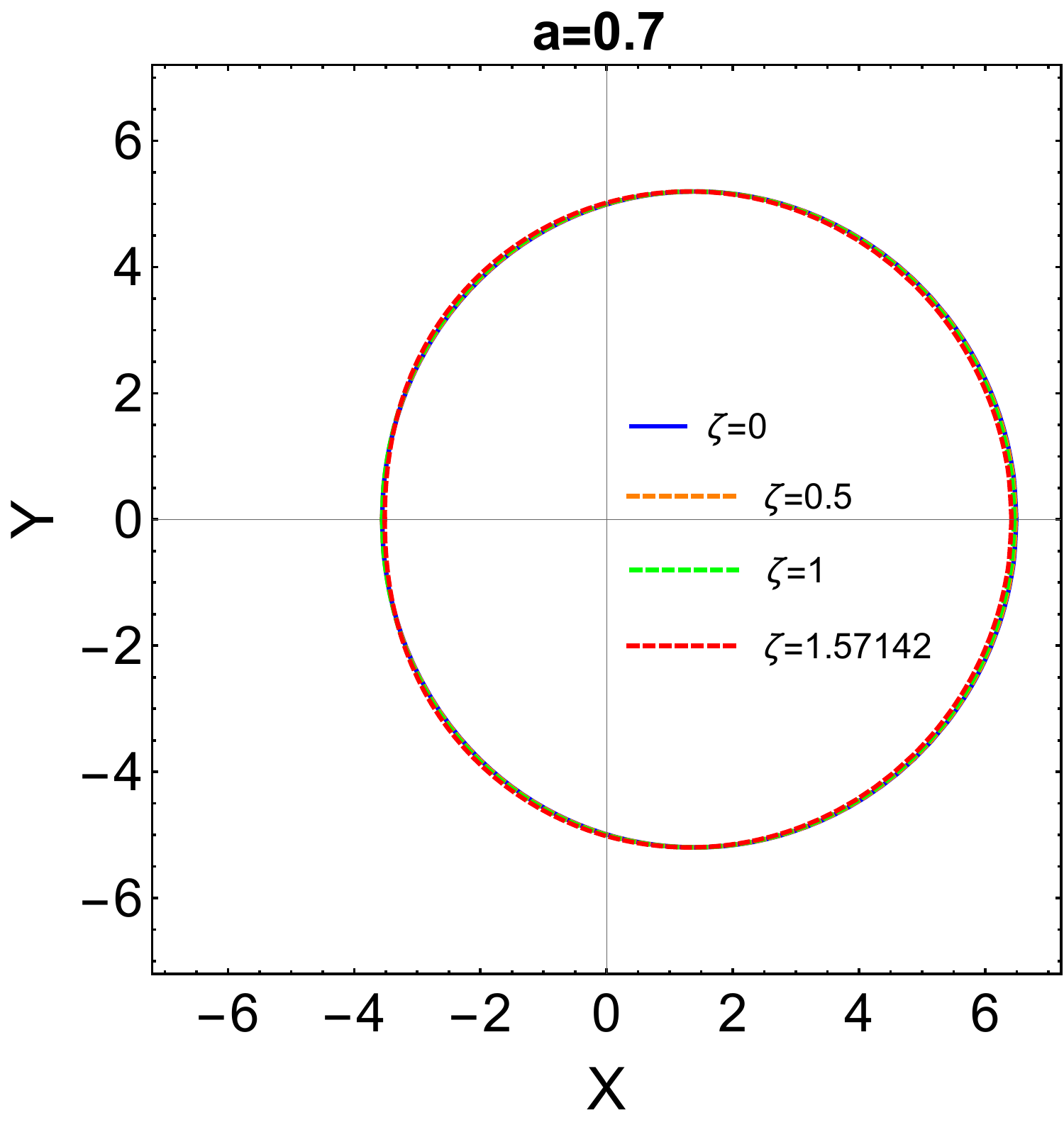}
	\end{minipage}
	\quad
	\begin{minipage}[t]{0.45\textwidth}
		\centering
		\includegraphics[width=3in, height=3in, keepaspectratio]{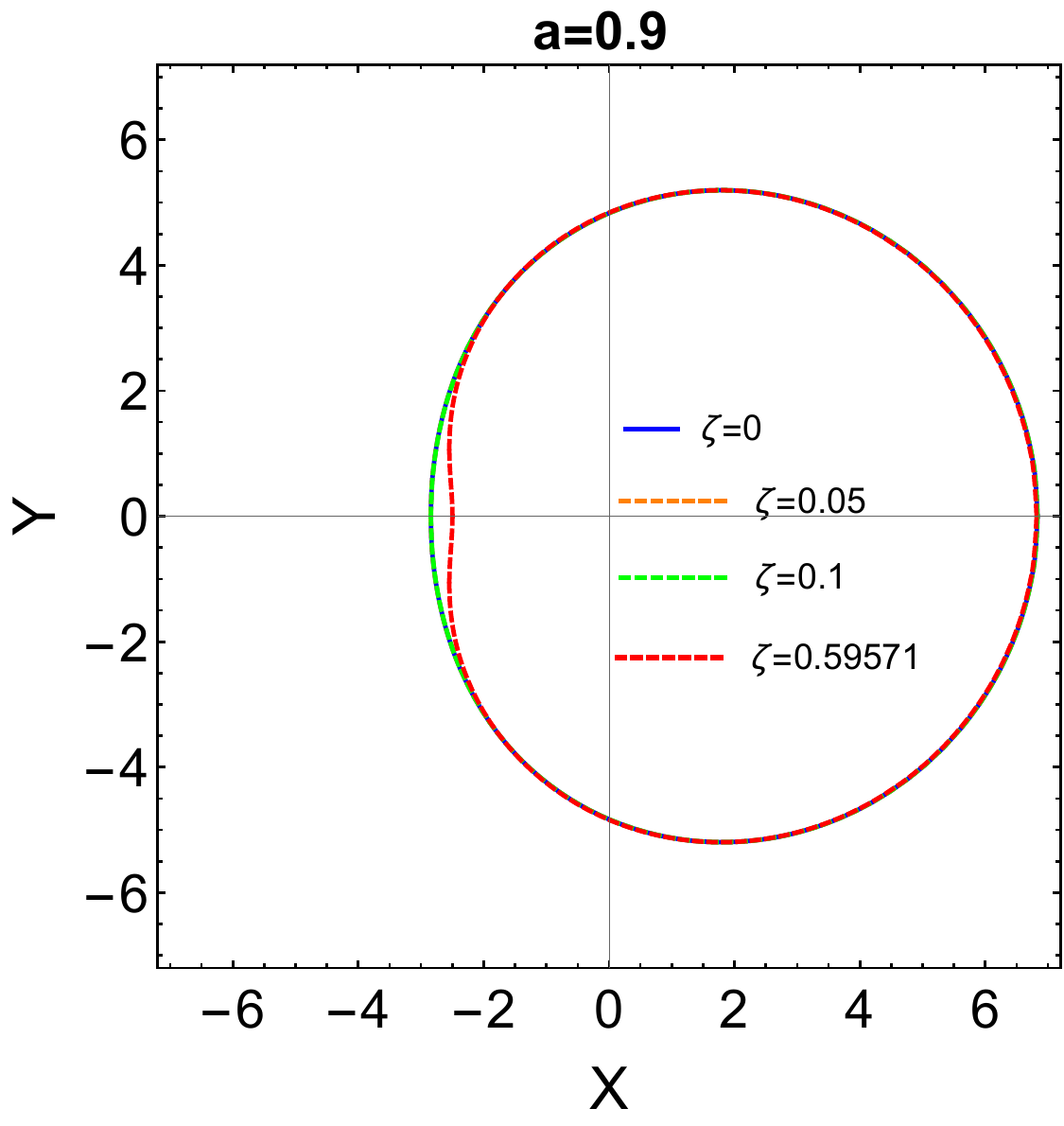}
	\end{minipage}
	\begin{minipage}[t]{0.45\textwidth}
		\centering
		\includegraphics[width=3in, height=3in, keepaspectratio]{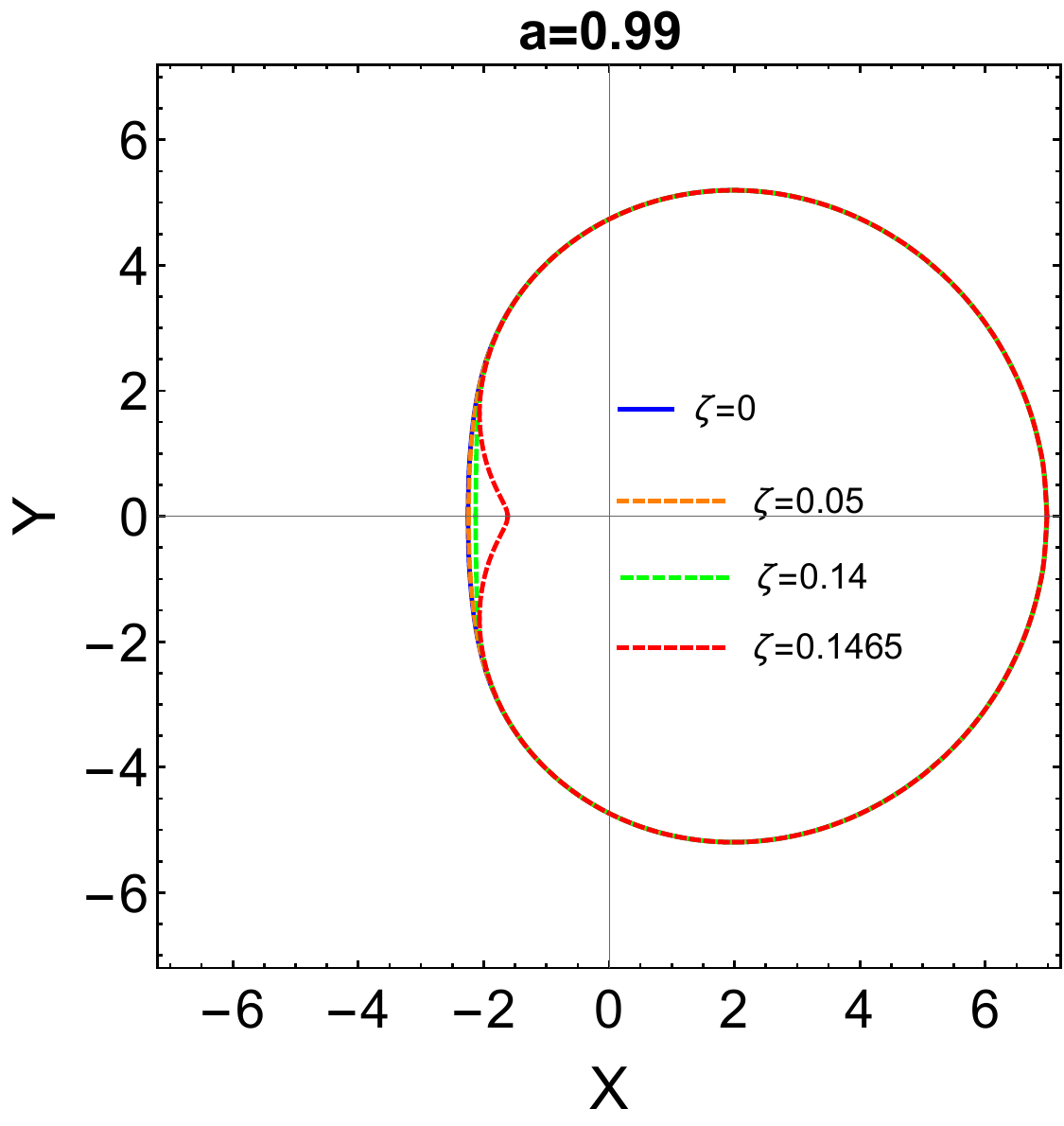}
	\end{minipage}
	\caption{Shadows cast by RBH-II for different parameters $a$ and $\zeta$.}
	\label{fig:rbh2}
\end{figure*}

With the allowed parameters, we plot the shadows cast by RBH-I and RBH-II and seen by a distant observer with $\theta_{\ast} = \frac{\pi}{2}$ in Figs. \ref{fig:rbh1} and \ref{fig:rbh2}, respectively. In these figures, the blue solid line represents the boundary of the shadow cast by a Kerr BH ($\zeta=0$), while the red dashed line indicates the one corresponding to the maximum allowed value of \(\zeta\) for a fixed \(a\). In each figure, the top two panels show the shadows in the non-extremal case, while the bottom two panels depict the ones in the near-extremal case. From Figs. \ref{fig:rbh1} and \ref{fig:rbh2}, we observe that the shadows of both rotating BHs share the same behavior with respect to $\zeta$ : (i) In the non-extremal case, the parameter $\zeta$ mainly affects the shadow size; (ii) In the near-extremal case, the parameter $\zeta$ will deform the shadow shape such that a cuspy edge arises. This provides a possible way to distinguish RBH-I and RBH-II from the Kerr BH by their shadows. By comparing Fig. \ref{fig:rbh1} with Fig. \ref{fig:rbh2}, we can see that the quantum parameter $\zeta$ has a significant influence on the size for the RBH-I shadow in the non-extremal case, and for the RBH-II shadow on the shape in the near-extremal case. Similarly, a cuspy edge also appears in the BH shadow cast by the Kerr BH with Proca hair \cite{Cunha:2017eoe}.

\begin{figure*}[htbp]
	\begin{minipage}[t]{0.45\textwidth}
		\centering
		\includegraphics[width=3in, height=3in, keepaspectratio]{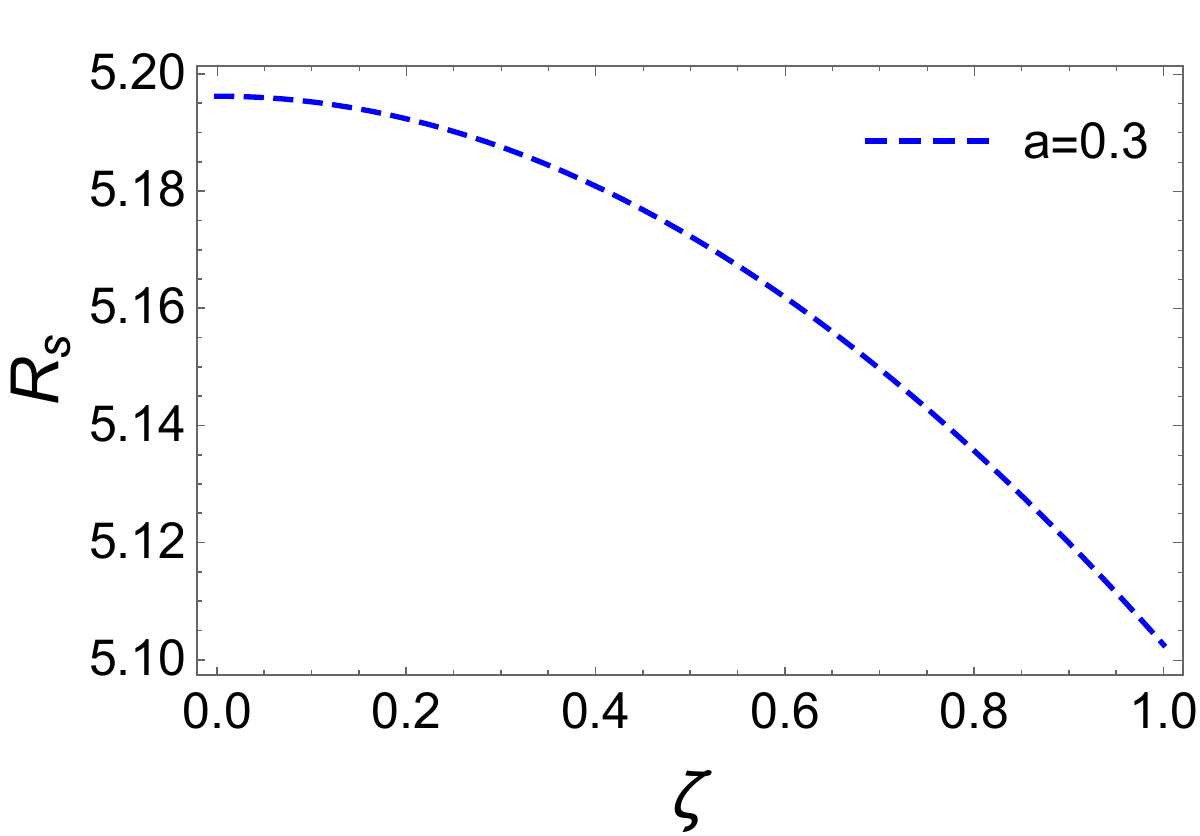}
	\end{minipage}
	\begin{minipage}[t]{0.45\textwidth}
		\centering
		\includegraphics[width=3in, height=3in, keepaspectratio]{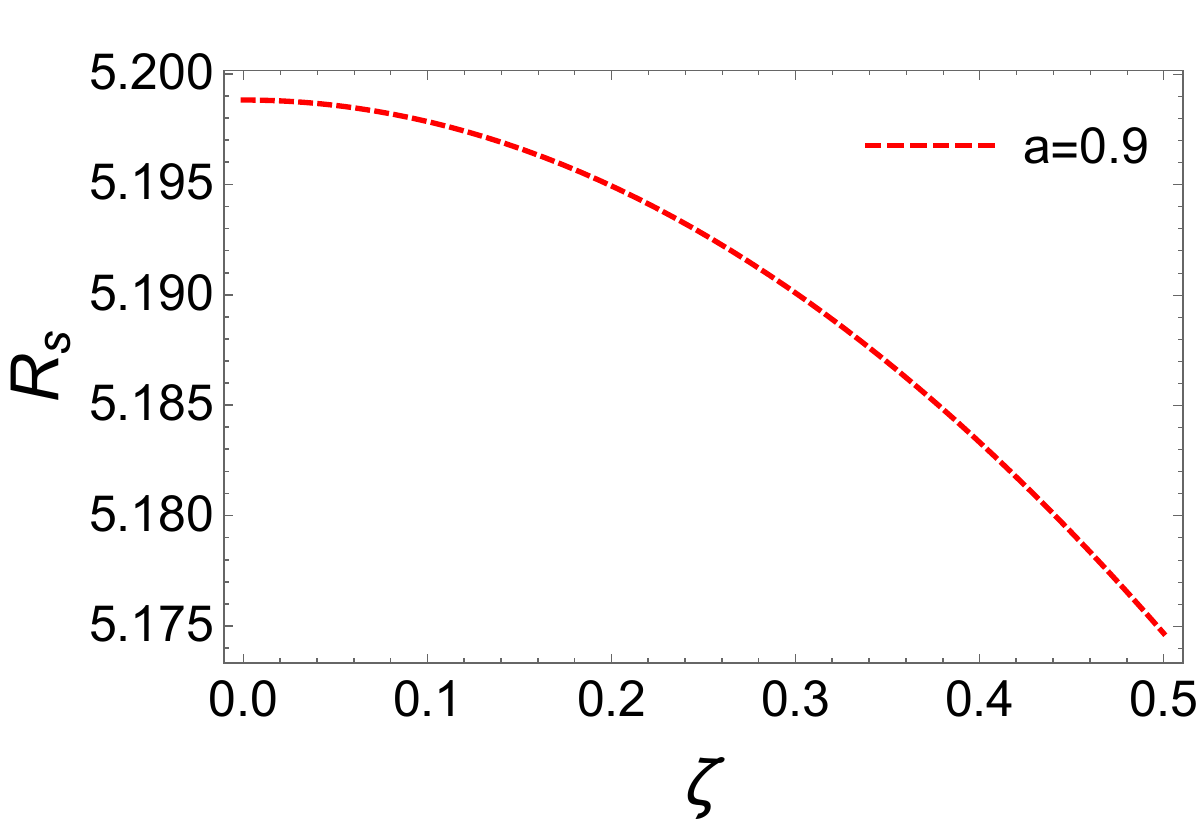}
	\end{minipage}
	\quad
	\begin{minipage}[t]{0.5\textwidth}
		\centering
		\includegraphics[width=3in, height=3in, keepaspectratio]{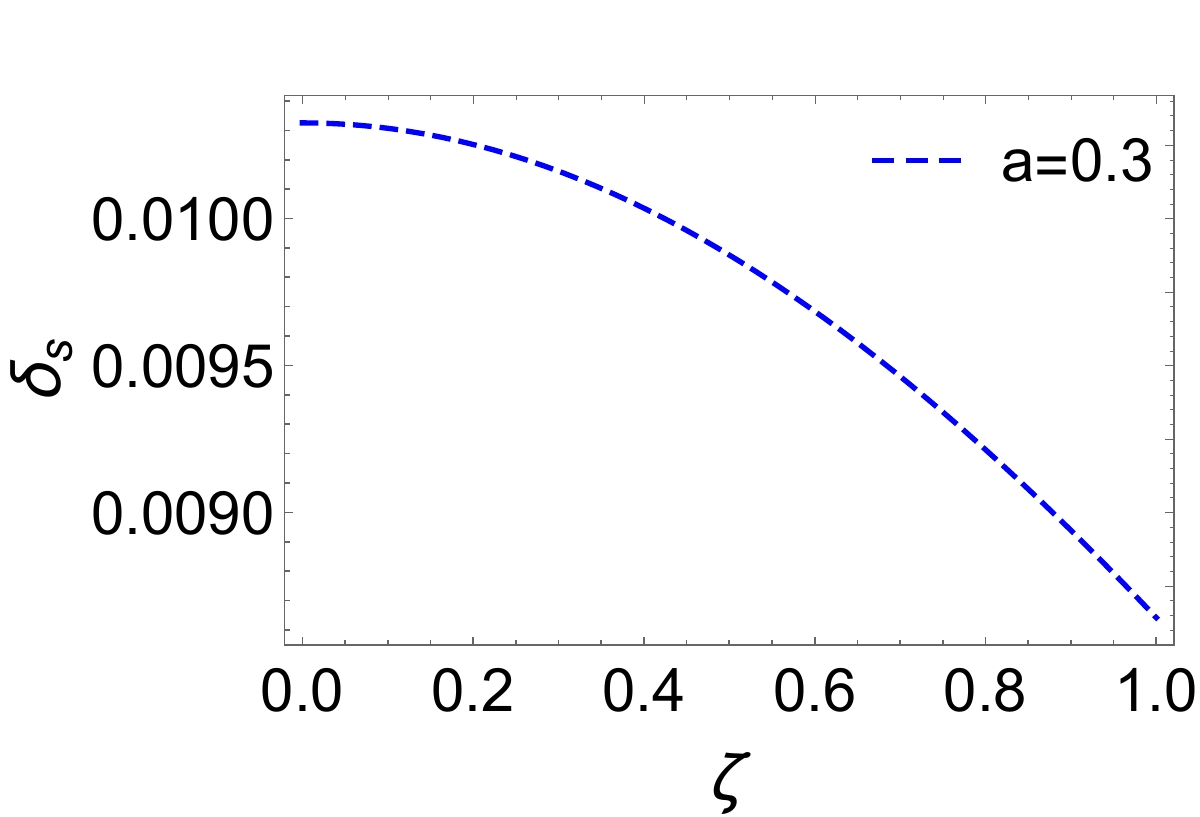}
	\end{minipage}
	\begin{minipage}[t]{0.45\textwidth}
		\centering
		\includegraphics[width=3in, height=3in, keepaspectratio]{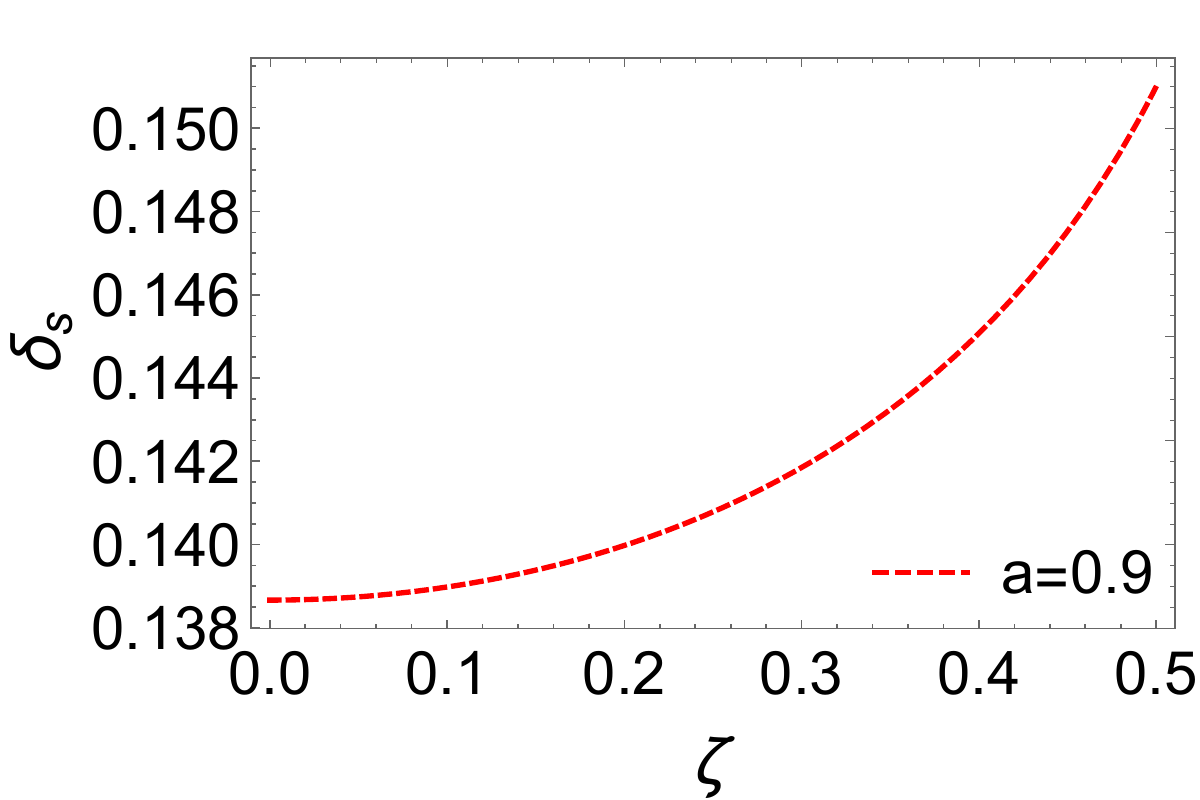}
	\end{minipage}
	\caption{The variation of \(R_s\) and \(\delta_s\) with \(\zeta\) for RBH-I: the parameter is \(a = 0.3\) in the left panel, and \(a = 0.9\) in the right panel.}
	\label{fig:rbh11}
\end{figure*}

\begin{figure*}[htbp]
	\begin{minipage}[t]{0.45\textwidth}
		\centering
		\includegraphics[width=3in, height=3in, keepaspectratio]{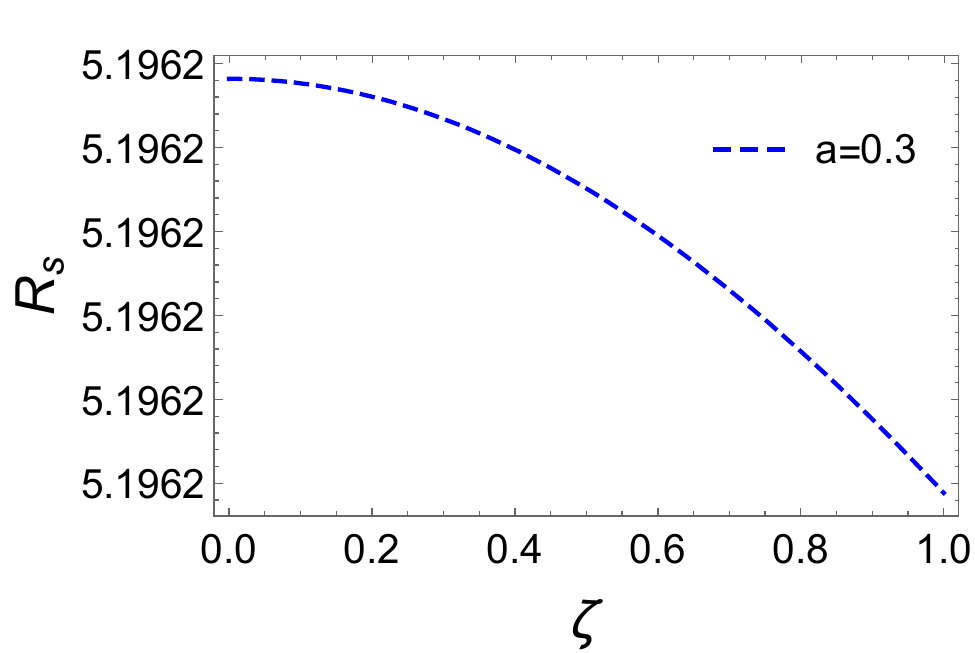}
	\end{minipage}
	\begin{minipage}[t]{0.45\textwidth}
		\centering
		\includegraphics[width=3in, height=3in, keepaspectratio]{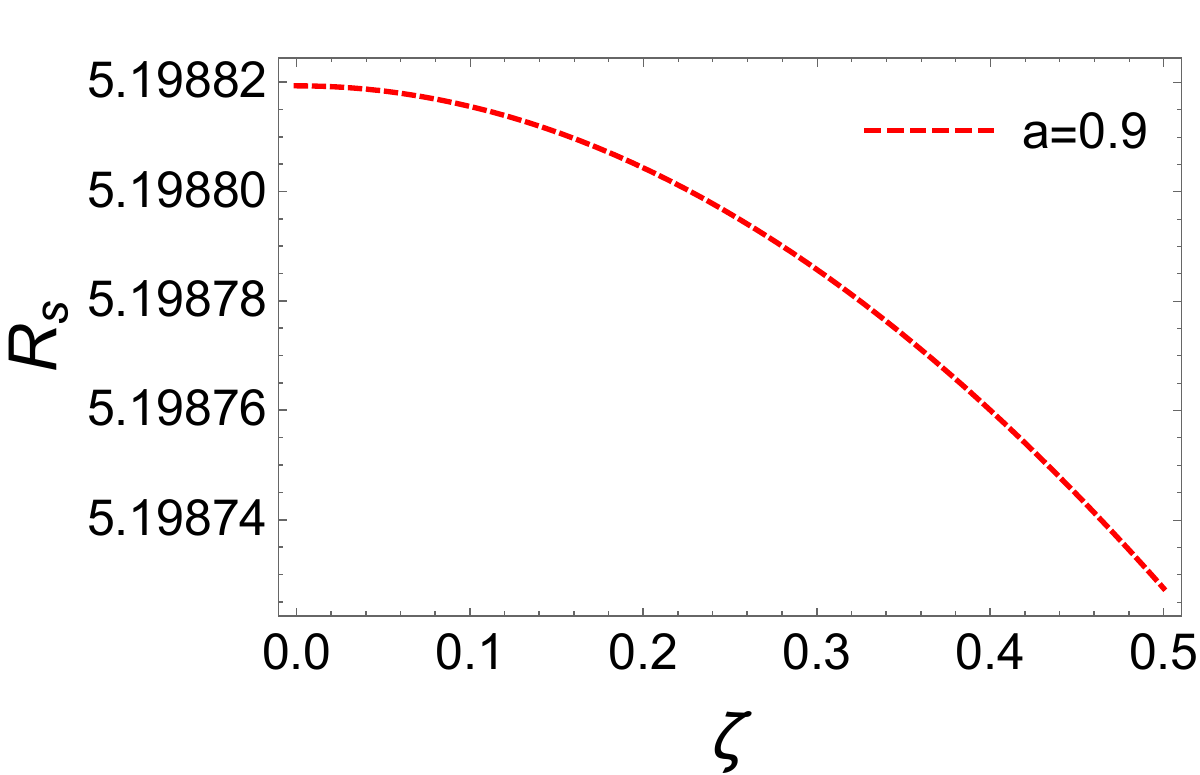}
	\end{minipage}
	\quad
	\begin{minipage}[t]{0.45\textwidth}
		\centering
		\includegraphics[width=3in, height=3in, keepaspectratio]{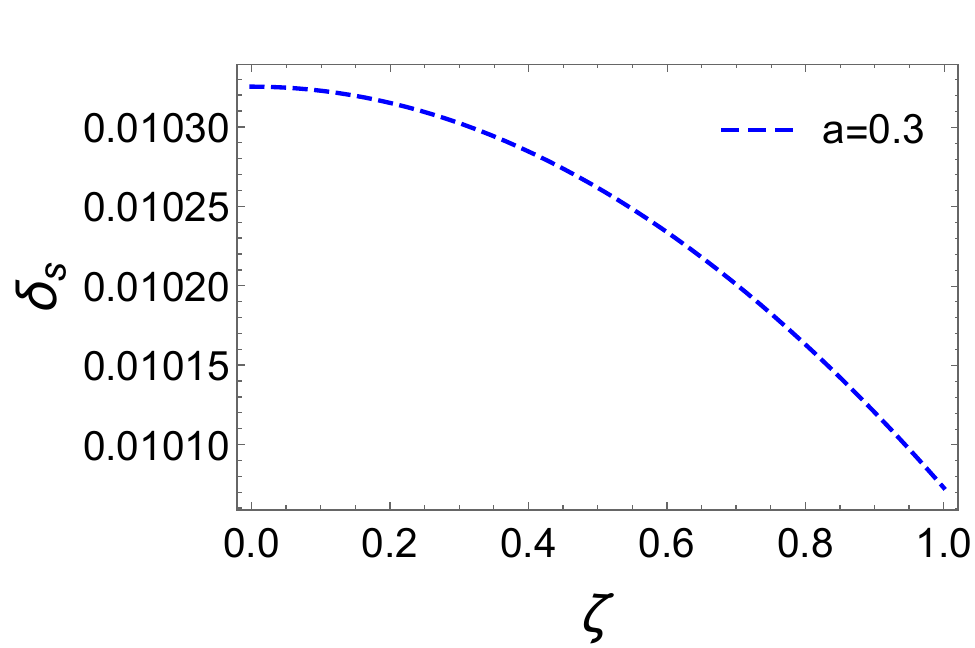}
	\end{minipage}
	\begin{minipage}[t]{0.45\textwidth}
		\centering
		\includegraphics[width=3in, height=3in, keepaspectratio]{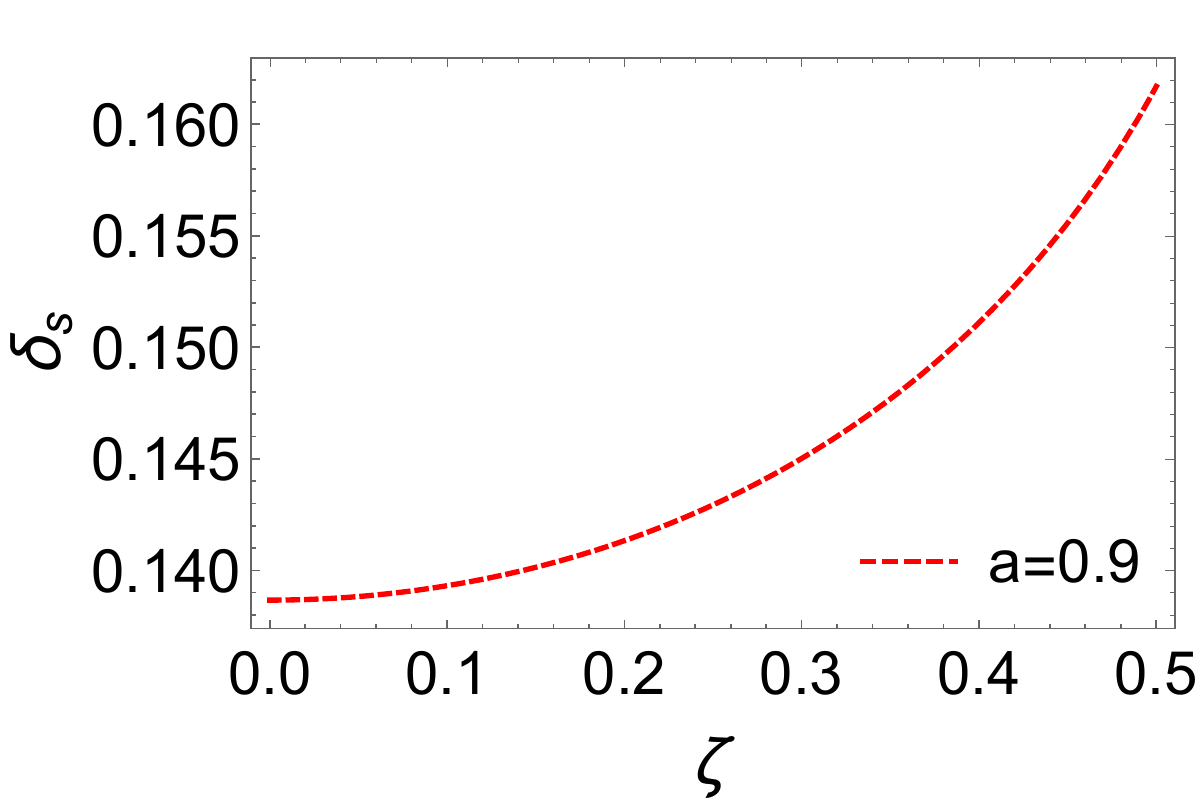}
	\end{minipage}
	\caption{The variation of \(R_s\) and \(\delta_s\) with \(\zeta\) for RBH-II: the parameter is \(a = 0.3\) in the left panel, and \(a = 0.9\) in the right panel.}
	\label{fig:rbh22}
\end{figure*}

To more precisely characterize the shadow feature, two observables related to the rotating BH shadow, the radius $R_s$ and the deformation $\delta_s$, are proposed in \cite{Hioki:2009na}. The radius $R_s$ is determined by three points on the shadow boundary: the top point $(x_t, y_t)$, the bottom one $(x_b, y_b)$, and the most right one $(x_r, 0)$, and is defined by \cite{Hioki:2009na}
\begin{equation}
	R_{s}:=\frac{\left(x_{r}-x_{t}\right)^{2}+y_{t}^{2}}{2\left(x_{r}-x_{t}\right)}.
\end{equation}
The deformation $\delta_s$ characterizing the shape deviation of the shadow relative to its standard circle reads \cite{Hioki:2009na}
\begin{align}
 \delta_s&:=\frac{D}{R_s}=\frac{\left|x_{l}-\tilde{x}_{l}\right|}{R_s},
\end{align}
where $D$ is the distance between the left vertex $(\tilde{x}_l, 0)$ of the standard circle with radius $R_s$ and the left vertex $(x_l, 0)$ of the actual shadow. The behaviors of \(R_s\) and \(\delta_s\) of the shadows cast by RBH-I and RBH-II were shown in Figs. \ref{fig:rbh11} and \ref{fig:rbh22}, respectively. These figures indicate the similar shadow behaviors with respect to $\zeta$ for both rotating BHs: (i) The presence of $\zeta$ shrinks the shadow size; (ii) The parameter $\zeta$ has a significant effect on shadow deformation in the near-extremal case with large $a$.

\section{Summary}\label{section5}
In this paper, we started from a brief review of two covariant spherically symmetric BH spacetimes \eqref{mt01} within effective quantum gravity, where the quantum gravity effect is encoded in the quantum parameter $\zeta$. Utilizing the modified NJA, we generated the rotating quantum-corrected BH solutions \eqref{xian} from the seed metrics \eqref{mt01}. These rotating metrics reduces to the Kerr metric as $\zeta\rightarrow0$.

To comprehend the effects of \(\zeta\) on the properties of the rotating BHs (RBH-I and RBH-II) and to compare the two BHs, we studied their horizons and static limit surfaces. Our results indicate that the horizon structures of RBH-I and RBH-II are the same. Varying the quantum parameter \(\zeta\) and the spin parameter \(a\) {leads to} different numbers of horizons as shown in Fig. \ref{fig:EH1}, with significant influence from \(\zeta\) only when \(a\) is large. Interestingly, we found that these rotating BHs can possess at most two horizons. Additionally, we investigated the behavior of the static limit surface and the outer horizon under various parameter values of \(\zeta\) and \(a\). Our findings reveal that both RBH-I and RBH-II exhibit similar properties in this context. For a fixed \(\zeta\), the sizes of the outer horizon and the static limit surface {decrease} as \(a\) increases. The static limit surface gradually approaches to the event horizon in the static case as \(a\) decreases. When \(a\) is fixed, the influences of \(\zeta\) on the outer horizon as well as on the static limit surface are very tiny. Hence, in comparison, the parameter \(a\) significantly affects the properties of these BHs.

By constraining $a$ and $\zeta$ in a reasonable range as shown in Fig. \ref{fig:a_zeta1}, we plotted the shadows cast by RBH-I and RBH-II in Figs. \ref{fig:rbh1} and \ref{fig:rbh2}, respectively. Our results show that for both rotating BHs, the parameter $\zeta$ mainly affects the shadow size in the non-extremal case, while it can deform the shadow shape by arising a cuspy edge in the near-extremal case. We also observed that RBH-I and RBH-II can be distinguished in the non-extremal case with small \(a\). By analyzing the impact of \(\zeta\) on the two observables \(R_s\) and \(\delta_s\) of the shadow, we found that for both rotating BHs, as \(\zeta\) increases, $R_s$ always decreases, and \(\delta_s\) decreases in the non-extremal case while increases in the near-extremal case.

\begin{acknowledgments}
This work is supported in part by NSFC Grants No. 12165005.
\end{acknowledgments}



\end{document}